\documentclass[journal]{IEEEtran}
\usepackage{graphicx,amsmath,url}
\usepackage{microtype,lmodern}
\pdfoutput=1 
\usepackage{mathtools,amssymb,gensymb}
\usepackage{tgtermes} 
\usepackage[T1]{fontenc}
\usepackage[utf8]{inputenc}
\usepackage{color,soul}
\usepackage[colorinlistoftodos]{todonotes}
\usepackage{acronym}
\usepackage{subcaption}
\usepackage{array}
\usepackage{booktabs}
\usepackage[compress]{cite}

\usepackage[normalem]{ulem}
\usepackage[switch]{lineno}
\usepackage{listings}

\usepackage{longtable}
\graphicspath{{figs/}}
\newcommand{\revised}[3]{#2}
\newcommand{\revisednolines}[1]{#1}

\definecolor{codegreen}{rgb}{0,0.6,0}
\definecolor{codegray}{rgb}{0.5,0.5,0.5}
\definecolor{codepurple}{rgb}{0.58,0,0.82}
\definecolor{backcolour}{rgb}{0.95,0.95,0.92}
 
\lstdefinestyle{mystyle}{basicstyle=\tiny\ttfamily,
  backgroundcolor=\color{backcolour},   
  commentstyle=\color{codegreen},
  keywordstyle=\color{magenta},
  numberstyle=\tiny\color{codegray},
  stringstyle=\color{codepurple},
  breakatwhitespace=false,         
  breaklines=true,                 
  captionpos=b,                    
  keepspaces=true,                 
  showspaces=false,                
  showstringspaces=false,
  showtabs=false,                  
  tabsize=2,
  frame=single,
  breaklines=true,
  breakatwhitespace=false
}

\lstdefinelanguage{chp}
{morekeywords={chp, hse, prs},
sensitive=false,
morecomment=[l]{//},
morecomment=[s]{/*}{*/},
morestring=[b]"
}

\begin{document}


%
\title{A scalable multi-core architecture with heterogeneous memory structures for Dynamic Neuromorphic Asynchronous Processors~(DYNAPs)}
%
%
%

\author{Saber~Moradi$^{\ast}$,~
        Ning~Qiao$^{\ddagger}$,~
        Fabio~Stefanini$^{\dagger}$,~
        and~Giacomo~Indiveri$^{\ddagger}$
       \\$^{\ast}$Computer Systems Lab, School of Engineering~\&\,Applied Sciences, Yale University, USA
       \\$^{\dagger}$Center of Theoretical Neuroscience, Columbia University, USA
       \\$^{\ddagger}$Institute of Neuroinformatics, University of Zurich and ETH Zurich, Switzerland
%
}

%
%

\markboth{}%
{Moradi \MakeLowercase{\textit{et al.}}: Bare Demo of IEEEtran.cls for IEEE Journals}
%

\acrodefplural{RAM}[RAMs]{Random Access Memories}
\acrodefplural{RRAM}[R-RAMs]{Resistive Random Access Memories}
\acrodefplural{STT-MRAM}[STT-MRAMs]{Spin-Transfer Torque Magnetic Random Access Memories}
\acrodef{ADC}[ADC]{Analog to Digital Converter}
\acrodef{ADEX}[AdExp-I\&F]{Adaptive-Exponential Integrate and Fire}
\acrodef{AER}[AER]{Address-Event Representation}
\acrodef{AEX}[AEX]{AER EXtension board}
\acrodef{AE}[AE]{Address-Event}
\acrodef{AFM}[AFM]{Atomic Force Microscope}
\acrodef{AGC}[AGC]{Automatic Gain Control}
\acrodef{AMDA}[AMDA]{AER Motherboard with D/A converters}
\acrodef{ANN}[ANN]{Attractor Neural Network}
\acrodef{API}[API]{Application Programming Interface}
\acrodef{ARM}[ARM]{Advanced RISC Machine}
\acrodef{ASIC}[ASIC]{Application Specific Integrated Circuit}
\acrodef{BCM}[BMC]{Bienenstock-Cooper-Munro}
\acrodef{BD}[BD]{Bundled Data}
\acrodef{BEOL}[BEOL]{Back-end of Line}
\acrodef{BG}[BG]{Bias Generator}
\acrodef{BMI}[BMI]{Brain-Machince Interface}
\acrodef{CAD}[CAD]{Computer Aided Design}
\acrodef{CAM}[CAM]{Content Addressable Memory}
\acrodef{CAVIAR}[CAVIAR]{Convolution AER Vision Architecture for Real-Time}
\acrodef{CFC}[CFC]{Current to Frequency Converter}
\acrodef{CCN}[CCN]{Cooperative and Competitive Network}
\acrodef{CHP}[CHP]{Communicating Hardware Processes}
\acrodef{CNN}[CNN]{Convolutional Neural Network}
\acrodef{CMIM}[CMIM]{Metal-insulator-metal Capacitor}
\acrodef{CMOL}[CMOL]{``Hybrid CMOS nanoelectronic circuits''}
\acrodef{CMOS}[CMOS]{Complementary Metal-Oxide-Semiconductor}
\acrodef{COTS}[COTS]{Commercial Off-The-Shelf}
\acrodef{CPG}[CPG]{Central Pattern Generator}
\acrodef{CPLD}[CPLD]{Complex Programmable Logic Device}
\acrodef{CPU}[CPU]{Central Processing Unit}
\acrodef{CV}[CV]{Coefficient of Variation}
\acrodef{DAC}[DAC]{Digital to Analog Converter}
\acrodef{DAS}[DAS]{Dynamic Auditory Sensor}
\acrodef{DAVIS}[DAVIS]{Dynamic and Active Pixel Vision Sensor}
\acrodef{DBN}[DBN]{Deep Belief Network}
\acrodef{DFA}[DFA]{Deterministic Finite Automaton}
\acrodef{DMA}[DMA]{Direct Memory Access}
\acrodef{DNF}[DNF]{Dynamic Neural Field}
\acrodef{DNN}[DNN]{Deep Neural Network}
\acrodef{DOF}[DOF]{Degrees of Freedom}
\acrodef{DPE}[DPE]{Dynamic Parameter Estimation}
\acrodef{DPI}[DPI]{Differential Pair Integrator}
\acrodef{DRAM}[DRAM]{Dynamic Random Access Memory}
\acrodef{DR}[DR]{Dual Rail}
\acrodef{DSP}[DSP]{Digital Signal Processor}
\acrodef{DVS}[DVS]{Dynamic Vision Sensor}
\acrodef{EBL}[EBL]{Electron Beam Lithography}
\acrodef{EDVAC}[EDVAC]{Electronic Discrete Variable Automatic Computer}
\acrodef{EIN}[EIN]{Excitatory-Inhibitory Network}
\acrodef{EM}[EM]{Expectation Maximization}
\acrodef{EPSC}[EPSC]{Excitatory Post-Synaptic Current}
\acrodef{EPSP}[EPSP]{Excitatory Post-Synaptic Potential}
\acrodef{FDSOI}[FDSOI]{Fully-Depleted Silicon on Insulator}
\acrodef{FET}[FET]{Field-Effect Transistor}
\acrodef{FFT}[FFT]{Fast Fourier Transform}
\acrodef{FI}[F-I]{Frequency-Current}
\acrodef{FPGA}[FPGA]{Field Programmable Gate Array}
\acrodef{FSA}[FSA]{Finite State Automaton}
\acrodef{FSM}[FSM]{Finite State Machine}
\acrodef{GOPS}[GOPS]{Giga-Operations per Second}
\acrodef{GPU}[GPU]{Graphical Processing Unit}
\acrodef{GUI}[GUI]{Graphical User Interface}
\acrodef{HAL}[HAL]{Hardware Abstraction Layer}
\acrodef{HH}[H\&H]{Hodgkin \& Huxley}
\acrodef{HMM}[HMM]{Hidden Markov Model}
\acrodef{HRS}[HRS]{High-Resistive State}
\acrodef{HR}[HR]{Human Readable}
\acrodef{HSE}[HSE]{Handshaking Expansion}
\acrodef{HW}[HW]{Hardware}
\acrodef{ICT}[ICT]{Information and Communication Technology}
\acrodef{IC}[IC]{Integrated Circuit}
\acrodef{IF2DWTA}[IF2DWTA]{Integrate \& Fire 2--Dimensional WTA}
\acrodef{IFSLWTA}[IFSLWTA]{Integrate \& Fire Stop Learning WTA}
\acrodef{IF}[I\&F]{Integrate-and-Fire}
\acrodef{IMU}[IMU]{Inertial Measurement Unit}
\acrodef{INCF}[INCF]{International Neuroinformatics Coordinating Facility}
\acrodef{INI}[INI]{Institute of Neuroinformatics}
\acrodef{IO}[I/O]{Input/Output}
\acrodef{IPSC}[IPSC]{Inhibitory Post-Synaptic Current}
\acrodef{IPSP}[IPSP]{Inhibitory Post-Synaptic Potential}
\acrodef{IP}[IP]{Intellectual Property}
\acrodef{ISI}[ISI]{Inter-Spike Interval}
\acrodef{JFLAP}[JFLAP]{Java - Formal Languages and Automata Package}
\acrodef{LLC}[LLC]{Low Leakage Cell}
\acrodef{LFP}[LFP]{Local Field Potential}
\acrodef{LNA}[LNA]{Low-Noise Amplifier}
\acrodef{LPF}[LPF]{Low-Pass Filter}
\acrodef{LRS}[LRS]{Low-Resistive State}
\acrodef{LSM}[LSM]{Liquid State Machine}
\acrodef{LTD}[LTD]{Long Term Depression}
\acrodef{LTI}[LTI]{Linear Time-Invariant}
\acrodef{LTP}[LTP]{Long Term Potentiation}
\acrodef{LTU}[LTU]{Linear Threshold Unit}
\acrodef{LUT}[LUT]{Look-Up Table}
\acrodef{MCMC}[MCMC]{Markov-Chain Monte Carlo}
\acrodef{MEMS}[MEMS]{Micro Electro Mechanical System}
\acrodef{MIM}[MIM]{Metal Insulator Metal}
\acrodef{MOSCAP}[MOSCAP]{Metal Oxide Semiconductor Capacitor}
\acrodef{MOSFET}[MOSFET]{Metal Oxide Semiconductor Field-Effect Transistor}
\acrodef{MOS}[MOS]{Metal Oxide Semiconductor}
\acrodef{MRI}[MRI]{Magnetic Resonance Imaging}
\acrodef{NDFSM}[NDFSM]{Non-deterministic Finite State Machine} 
\acrodef{ND}[ND]{Noise-Driven}
\acrodef{NEF}[NEF]{Neural Engineering Framework}
\acrodef{NHML}[NHML]{Neuromorphic Hardware Mark-up Language}
\acrodef{NIL}[NIL]{Nano-Imprint Lithography}
\acrodef{NMDA}[NMDA]{N-Methyl-D-Aspartate}
\acrodef{NME}[NE]{Neuromorphic Engineering}
\acrodef{OTA}[OTA]{Operational Transconductance Amplifier}
\acrodef{PCB}[PCB]{Printed Circuit Board}
\acrodef{PFM}[PFM]{Pulse Frequency Modulation}
\acrodef{PR}[PR]{Production Rule}
\acrodef{PSC}[PSC]{Post-Synaptic Current}
\acrodef{PSTH}[PSTH]{Peri-Stimulus Time Histogram}
\acrodef{QDI}[QDI]{Quasi Delay Insensitive}
\acrodef{RAM}[RAM]{Random Access Memory}
\acrodef{RMSE}[RMSE]{Root Mean Squared-Error}
\acrodef{RMS}[RMS]{Root Mean Squared}
\acrodef{RNN}[RNN]{Recurrent Neural Network}
\acrodef{ROLLS}[ROLLS]{Reconfigurable On-Line Learning Spiking}
\acrodef{RRAM}[RRAM]{Resistive Random Access Memory}
\acrodef{SAC}[SAC]{Selective Attention Chip}
\acrodef{SCX}[SCX]{Silicon CorteX}
\acrodef{SD}[SD]{Signal-Driven}
\acrodef{SEM}[SEM]{Spike-based Expectation Maximization}
\acrodef{SLAM}[SLAM]{Simultaneous Localization and Mapping}
\acrodef{SOC}[SOC]{System-On-Chip}
\acrodef{SOI}[SOI]{Silicon on Insulator}
\acrodef{SRAM}[SRAM]{Static Random Access Memory}
\acrodef{STDP}[STDP]{Spike-Timing Dependent Plasticity}
\acrodef{STD}[STD]{Short-Term Depression}
\acrodef{STP}[STP]{Short-Term Plasticity}
\acrodef{STT-MRAM}[STT-MRAM]{Spin-Transfer Torque Magnetic Random Access Memory}
\acrodef{STT}[STT]{Spin-Transfer Torque}
\acrodef{SW}[SW]{Software}
\acrodef{TFT}[TFT]{Thin Film Transistor}
\acrodef{USB}[USB]{Universal Serial Bus}
\acrodef{VHDL}[VHDL]{VHSIC Hardware Description Language}
\acrodef{VLSI}[VLSI]{Very Large Scale Integration}
\acrodef{VOR}[VOR]{Vestibulo-Ocular Reflex}
\acrodef{WTA}[WTA]{Winner-Take-All}
\acrodef{XML}[XML]{eXtensible Mark-up Language}
\acrodef{divmod3}[DIVMOD3]{divisibility of a number by 3}
\acrodef{hWTA}[hWTA]{Hard Winner-Take-All}
\acrodef{sWTA}[sWTA]{soft Winner-Take-All}

\maketitle

\begin{abstract}
Neuromorphic computing systems comprise networks of neurons that use asynchronous events for both computation and communication. This type of representation offers several advantages in terms of bandwidth and power consumption in neuromorphic electronic systems. However, managing the traffic of asynchronous events in large scale systems is a daunting task, both in terms of circuit complexity and memory requirements.
Here we present a novel routing methodology that employs both hierarchical and mesh routing strategies and combines heterogeneous memory structures for minimizing both memory requirements and latency, while maximizing programming flexibility to support a wide range of event-based neural network architectures, through parameter configuration. We validated the proposed scheme in a prototype multi-core neuromorphic processor chip that employs hybrid analog/digital circuits for emulating synapse and neuron dynamics together with asynchronous digital circuits for managing the address-event traffic. We present a theoretical analysis of the proposed connectivity scheme, describe the methods and circuits used to implement such scheme, and characterize the prototype chip. Finally, we demonstrate the use of the neuromorphic processor with a convolutional neural network for the real-time classification of visual symbols being flashed to a dynamic vision sensor~(DVS) at high speed.
\end{abstract}
\begin{IEEEkeywords}
Neuromorphic computing, routing architectures, asynchronous, circuits and systems
\end{IEEEkeywords}



%

\section{Introduction} %
\label{sec:introduction}
In an effort to develop a new generation of brain-inspired non von Neumann computing systems, several neuromorphic computing platforms have been proposed in recent years, that implement spike-based re-configurable neural networks~\cite{Moradi_Indiveri14, Qiao_etal15, Benjamin_etal14, Merolla_etal14a, Furber_etal14,Scholze_etal12, Brink_etal13, Moradi_Indiveri11}. Despite being developed with different goals in mind and following different design strategies, \revised{aer-cirrection}{most of}{all} these architectures share the same data representation and signal communication protocol: the \ac{AER}~\cite{Deiss_etal94,Boahen00}. In this representation computational units (e.g., neurons) are assigned an address that is encoded as a digital word and transmitted as soon they produce an event (e.g., as soon as  the neuron spikes) using asynchronous digital circuits.  Information is therefore encoded in the timing of these address-events. In event-based neural networks the neurons inter-spike interval (the interval between successive address events produced by the same neuron) represent the analog data, and neural computation is achieved by connecting multiple neurons among each other with different types of connectivity schemes. Spike produced by source neurons are transmitted to one or more destination synapse circuits that integrate them with different gain factors and convey them to the post-synaptic neuron. Unlike classical digital logic circuits, these networks are typically characterized by very large fan-in and fan-out numbers. For example, in cortical networks neurons project on average to about 10000 destinations. The type of processing and functionality of these spiking neural networks is determined \revised{rev1-1}{}{the} by their specific structure and parameters, such as the properties of the neurons or the weights of the synapses~\cite{Dayan_Abbott01}. It is therefore important to design neuromorphic computing platforms that can be configured to support the construction of different network topologies, with different neuron and synapse properties. This requires the development of configurable neuron and synapse circuits and of programmable \ac{AER} routing and communication schemes. The latter elements are particularly important, because the scalability of neuromorphic systems is mainly restricted by communication requirements. Indeed, some of the main bottlenecks in the construction of large-scale re-configurable neuromorphic computing platforms are the bandwidth, latency, and memory requirements for routing address-events among neurons. 
Most large-scale neuromorphic computing approaches followed up to now have either restricted the space of possible network connectivity schemes to optimize bandwidth usage while minimizing power and latency~\cite{Benjamin_etal14,Zamarreno-Ramos_etal13}, or have designed systems that use large amounts of memory, silicon real-estate, and/or power, to maximize flexibility and programmability~\cite{Merolla_etal14a,Furber_etal14}. In particular, most approaches proposed either make use of 2D mesh routing schemes, with maximum flexibility, but at the cost of large resource usage, or tree routing schemes which minimize latency and power, but are more restrictive in the types of networks that can be supported (see~\cite{Park_etal16} for a comprehensive overview comparing most of the approaches that have been proposed in the literature). In~\cite{Park_etal16} the authors proposed a \emph{hierarchical} address event routing scheme (HiAER) that overcomes some of the limitations of previous flat tree-based approaches. However, as with many of the previous approaches~\cite{Furber_etal14}, the HiAER architecture stores the routing tables in external memory banks implemented using \ac{DRAM}. Within this context, these approaches do not represent a radical departure from the classical von Neumann architecture, as they are affected by the von Neumann bottleneck problem~\cite{Backus78,Indiveri_Liu15}.

In this paper we propose a radically new routing scheme that integrates within the arrays of neurons and synapses both asynchronous routing circuits and heterogeneous memory structures used to store the data as distributed programmable routing tables. The routing scheme has been obtained by analyzing all previous approaches, and carrying out a systematic study for minimizing memory resources and maximizing network programmability and flexibility. The approach we propose combines the advantages of all previous approaches proposed up to now by 
combining 2D-mesh with hierarchical routing, implementing a 2-stage routing scheme for minimizing memory usage, which employs a combination of point-to-point source-address routing and multi-cast destination-address routing, and by using heterogeneous memory structures distributed within and across the neuron and synapse arrays which are optimally suited to exploit emerging memory technologies such as \ac{RRAM}~\cite{Akinaga_Shima10}.

In the Section~\ref{sec:optimized} we present the theory that was developed to minimize memory requirements in multi-neuron architectures. In Section~\ref{sec:hierarchical} we describe the mixed mode hierarchical-mesh routing scheme proposed, and the \ac{QDI} circuits designed to implement them; in Section~\ref{sec:hardware} we present a multi-core neuromorphic chip that was designed and fabricated to validate this routing scheme.  The chip comprises 1k VLSI neurons distributed among four neural cores and programmable routers for managing the \ac{AER} traffic. The routing fabric implemented in this chip uses new designs of quasi-delay insensitive asynchronous circuits, synthesized following the Communicating Hardware Process (CHP) formalism, while the neural and synaptic dynamics are designed using ultra-low power subthreshold neuromorphic circuits~\cite{Chicca_etal14}. Finally, in Section~\ref{sec:experimental-results} we demonstrate an application of this architecture by configuring the neuromorphic processor to implement a \ac{CNN} designed to process fast sequences of visual patterns, represented by \acp{AE} that are generated by a \ac{DVS} for low latency classification. We conclude the paper by comparing the features of the circuits and approach proposed to the current state-of-the-art and discuss the benefits of the proposed architecture.

\section{Memory optimized routing}
\label{sec:optimized}
Consider a generic spiking neural network with $N$ neurons in which each neuron has a large fan-out of $F$. 
This is the case for many neural network models including biologically plausible models of cortical networks, recurrent neural networks, convolutional networks and deep networks.
In standard routing schemes \revised{routingsd}{e.g., source or destination-based methods}{}, each neuron would be assigned a unique address encoded with $\log_2(N)$ bits. Therefore, to support $F$ fan-out destinations per neuron, a storage of $F\log_2(N)$ bits/neuron would be required. In total, the size of the connection memory the whole network would then be $NF\log_2(N)$.
While this scheme can support any type of connectivity, it is very wasteful in case of networks that have a specific structure, i.e., for most of the real and artificial neural networks~\cite{Litwin-Kumar_Doiron14,LeCun_etal15}.

\begin{figure}
  \centering
  \includegraphics[width=0.38\textwidth]{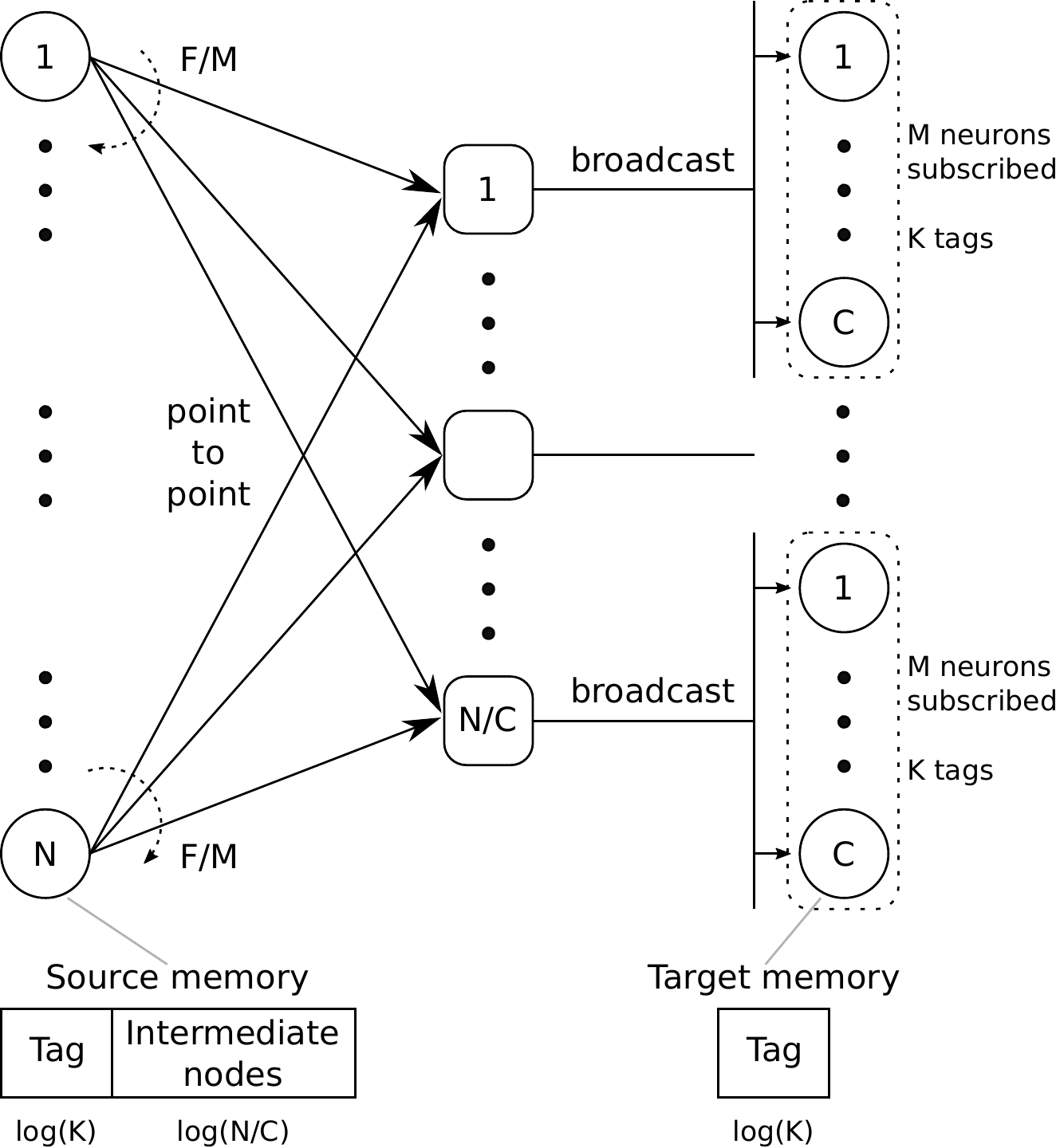}
  \caption{Two-stage tag-based routing scheme. The connectivity of a network of $N$ neurons with fan-out $F$ is implemented by a scheme that uses $N$ neurons with a reduced fan-out of $F/M$ that transmit their tag data to $N/C$ intermediate nodes, with point-to-point routing. Each intermediate node broadcasts its tag data to all neurons of its target cluster. Each cluster has $C$ neurons, with a subset of $M$ neurons subscribed to the incoming tag. The total unique number of tags used in each cluster is $K$. \revisednolines{Note that the nodes  on the right side represent the same neurons of the left side, but grouped into $N/C$ clusters.}{} }
  \label{fig:2-stage}
\end{figure}
Inspired by the connectivity patterns in biological neural networks~\cite{Douglas_Martin04,Muir_etal11} we developed a novel routing scheme
\revised{r4-R23}{that exploits their clustered connectivity structure to reduce memory requirements in large-scale neural networks. This novel method uses a mixed tag-based shared-addressing scheme, instead of  plain source or destination addressing schemes. The main idea is to introduce different clusters with independent address spaces, so that it is possible to re-use the same tag ids for connecting neurons among each other, without loss of generality. Specifically, in the proposed routing method neurons are  grouped in separate $N/C$ clusters of size $C$, and the connectivity with destination nodes (i.e., the fan-out operation)}{in which neurons are  grouped in $N/C$ clusters of size $C$, and the fan-out operation} is divided into two distinct stages~\cite{Moradi_etal13}.
In the first stage, the neurons use source-address routing for targeting \revised{parameters}{a subset~($F/M$, with $F>M$) of the intermediate nodes. The number of intermediate nodes is $N/C$ (with $F/M \leqslant N/C$).}{($F/M$)of the intermediate nodes} In the second stage the intermediate nodes \revised{rev4-R27}{broadcast the incoming events}{use destination-address routing} for targeting a number $M\leq C$ of destination neurons within each local domain (or cluster) $C$ (see Fig.~\ref{fig:2-stage}). \revised{redundant}{}{In particular, the intermediate nodes broadcast their \ac{AER} packet to all $C$  neurons in their end-point cluster.} Each neuron in the end-point cluster uses a set of \emph{tags} (one of $K$ tags for the cluster) to determine whether to accept or ignore the \ac{AER} packet received. In addition to distributing the memory resources among multiple  nodes (neurons)  within the computing fabric, therefore eliminating the von Neumann bottleneck~\cite{Backus78,Indiveri_Liu15}, this two-stage routing scheme allows us to minimize the total digital memory for configuring the network connectivity\footnote{provided that $M \leq F$ and $M\leq C$ -- see Appendix~\ref{sec:rout-memory-minim} for an in-depth analysis of the feasibility of these constraints.}, still allowing for large $M$-way fan-out: the total memory $MEM$ required per neuron can be separated into \emph{Source~memory} $MEM_S$ and \emph{Target~memory} $MEM_T$. At the sender side each source neuron stores $F/M$ entries. Each of these entries has two fields: the first field requires $\log_2(K)$ bits for representing a source tag and the second requires $\log_2(N/C$) bits to represent the address of the target intermediate nodes (see also Fig.~\ref{fig:2-stage}). Therefore the total \emph{Source~memory} required is $MEM_S = (F/M)(\log_2(K) + \log_2(N/C))$ bits/neuron.
At the destination side, assuming that the tags are uniformly distributed, each neuron needs to store $M$ tags. This leads to using  $KM$ tag entries per cluster and $KM/C$ tags per neuron, with each tag requiring $\log_2(K)$ bits. As a consequence, a total \emph{Target~memory} of $MEM_T = (KM/C)(\log_2(K))$ bits/neuron is required. Taking into account these values, we get: 
\begin{align}
  \label{eq:bits-neuron-1}
  MEM &= MEM_S + MEM_T 
\end{align}   

\begin{align}
  \label{eq:bits-neuron}
  MEM &= \frac{F}{M}\log_2 \left( {\frac{KN}{C}} \right) + \frac{KM}{C}\log_2(K) 
\end{align}   

As evident from eq.~\ref{eq:bits-neuron}, larger clusters and fewer tags lead to reduced memory requirements.  Reducing the number of tags  reduces the routing flexibility. However this can be counter balanced by increasing the number of clusters. By considering the ratio $\alpha=K/C$ it is possible to minimize for memory usage while maximizing for routing flexibility: if we substitute $K=\alpha C$ in eq.\ref{eq:bits-neuron}, we get:
\begin{align}
\label{eq:bits-flat}
  MEM &= \frac{F}{M}\log_2(\alpha N) + \alpha M\log_2(\alpha C)
\end{align}
The parameter $M$ determines the trade-off between point-to-point copying in the first stage versus broadcasting in the second stage of the routing scheme. The total memory requirement can therefore be minimized by differentiating eq.~\ref{eq:bits-flat} with respect to $M$, and determining the optimal $M^*$:
\begin{align}
  0 & =  \alpha\log_2(\alpha C) - \frac{F}{M^{2}}\log_2(\alpha N) \\
  M^* & =  \sqrt{\frac{F}{\alpha}\frac{\log_2(\alpha N)}{\log_2(\alpha C)} }
\end{align}
With this choice of $M$, the total number of bits per neuron required are $MEM = 2\sqrt{\alpha F\log_2(\alpha C)\log_2(\alpha N)}$.
If for example we set, as a design choice, $\alpha=1$, then we obtain $M^* = \sqrt{F\log_2(N)/\log_2(C)}$.
This leads to a total memory/neuron requirement of:
\begin{align}
MEM & = 2\sqrt{F\log_2(C)\log_2(N)} 
\end{align}
We therefore developed a scheme in which the memory requirements scale with $N$ in a way that is drastically lower than the scaling of standard routing schemes, that require $F\log_2(N)$ bits/neuron (e.g.\ compare 160k\,bits/neuron of memory required for a network of approximately 1 million ($2^{20}$) neurons, with fan-out of almost 10000 ($2^{13}$) with the conventional routing approach, versus less than 1.2k\,bits/neuron required for our memory-optimized scheme, for a network of same size, with same fan-out, and with cluster size of 256 neurons).\\
\revised{add_routing}{Unlike basic source or destination based routing methods, the proposed scheme is a mixed one that makes use of tag addresses which are shared among source and destination neurons. This results in a smaller address space to encode the connections among neurons and fewer memory entries to support large fan-out in biologically plausible clustered neural networks.}{}

\section{Mixed-mode hierarchical-mesh routing architecture}
\label{sec:hierarchical}
\begin{figure*}
  \centering
  \includegraphics[width=0.48\textwidth]{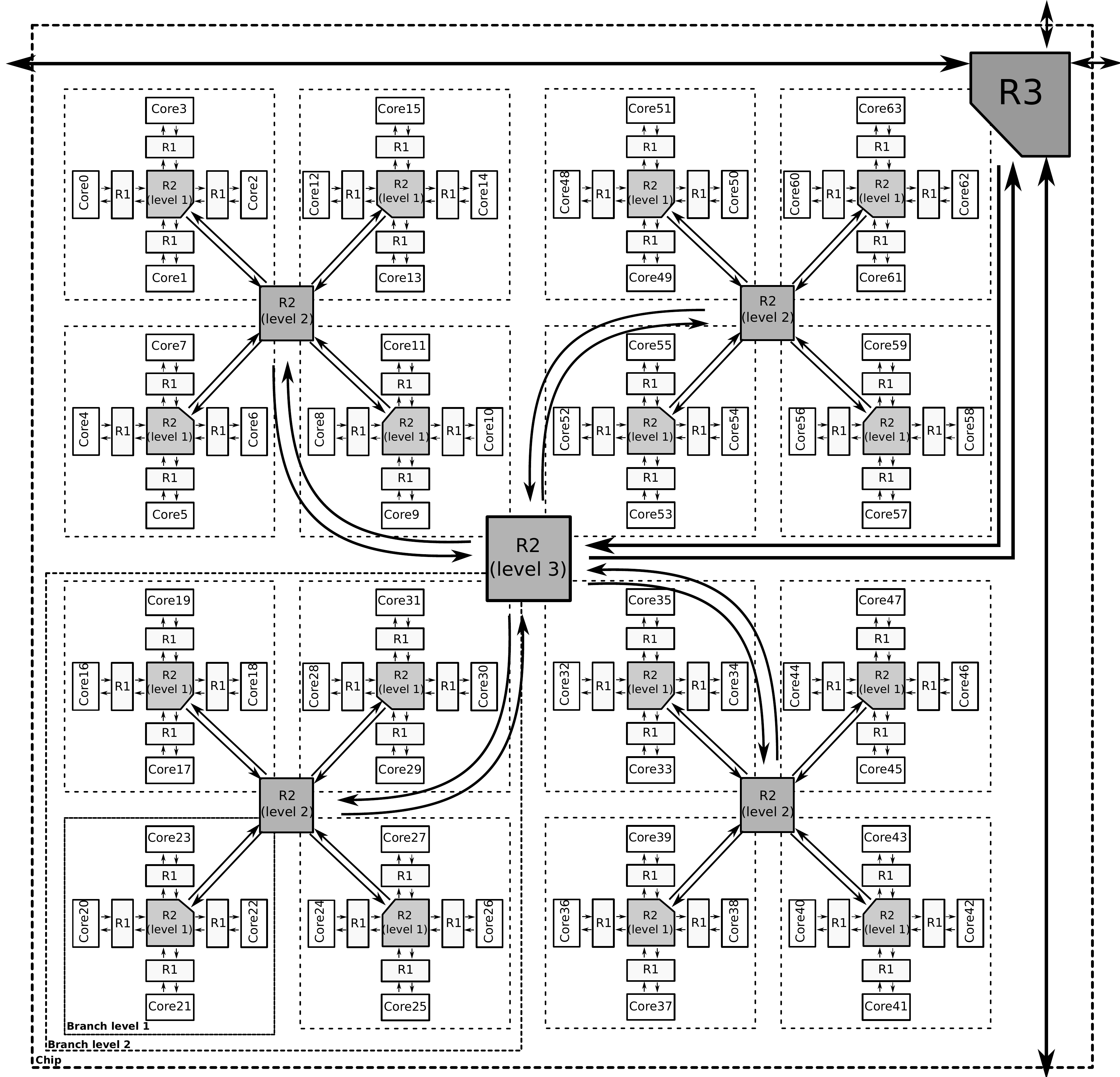}
  \caption{Mixed-mode hierarchical-mesh routing network example. At the lowest level, single cores transmit and receive events via broadcast operations through a ``R1'' router. Groups of four cores are linked together by a level-1 ``R2'' router. To reach cores belonging to different groups (but for example on the same tile), it is possible to use ``level-2'' or higher level R2 routers, following a tree-based routing strategy. To reach even further destinations (e.g.\ on different tiles), it is possible to use an ``R3'' router, which transmits signals along four cardinal directions using a 2D-mesh routing strategy.}
  \label{fig:routers}
\end{figure*}

In this section we propose a multi-core routing architecture that deploys the memory optimization scheme of Section~\ref{sec:optimized}: each cluster of Fig.~\ref{fig:2-stage} is mapped onto a ``core'',  the intermediate nodes are implemented by asynchronous routers, and the tags are stored into asynchronous \ac{CAM} blocks. The \ac{CAM} block (of each neuron) contains multiple tags representing the address of sources that this neuron is subscribed to. 

To optimize the event routing within the network, we adopted a mixed-mode approach that combines the advantages of mesh routing schemes \revised{bandwidth1}{(low bandwidth requirements, but high latency)}{(low bandwidth, but high latency)}, with those of hierarchical routing ones \revised{bandwidth2}{(low latency, but high bandwidth requirements)}{(low latency, but low bandwidth)}~\cite{Benjamin_etal14}.
Specifically, this architecture adopts a hierarchical routing scheme with three distinct levels of routers: at the lowest level an R1 is responsible for the local traffic, it either sends back the events to the present core or the next level of hierarchy. The events sent from the R1 to the core are  broadcast to all nodes within the core; following the prescription of the scheme described in Section~\ref{sec:optimized}, In other words, incoming events are presented to the \ac{CAM} blocks of all neurons and consequently they will be accepted by the nodes as valid inputs if there is match in the \ac{CAM}. The non-local events are sent to the second level router (R2) which has bidirectional channels to communicate with local cores and the next level router. Depending on the complexity of the network, this tree-based routing structure can have multiple R2 levels (e.g., see Fig.~\ref{fig:routers} for an example with three R2 levels). For transmitting data packets across even longer distances, the highest level router (R3) is employed, which uses a relative 2D-mesh~(also known as xy algorithm) routing strategy.
A combination of R1, R2, and R3 routers represent the intermediate nodes of Fig.~\ref{fig:2-stage}, the cores represent the clusters, the \emph{Source~memory} is stored in \ac{SRAM} cells in the R1 routers, and the \emph{Target~memory} in the synapse \ac{CAM} cells.
For the prototype described in this paper, we have chosen three level of routing. The routing network in our implementations is based on the use of ultra-low power and low-latency \ac{QDI} asynchronous digital routing and communication circuits.

\subsection{Quasi Delay-Insensitive asynchronous circuit design methodology}
\label{sec:QDI}

\begin{figure}
  \centering
  \includegraphics[width=0.47\textwidth]{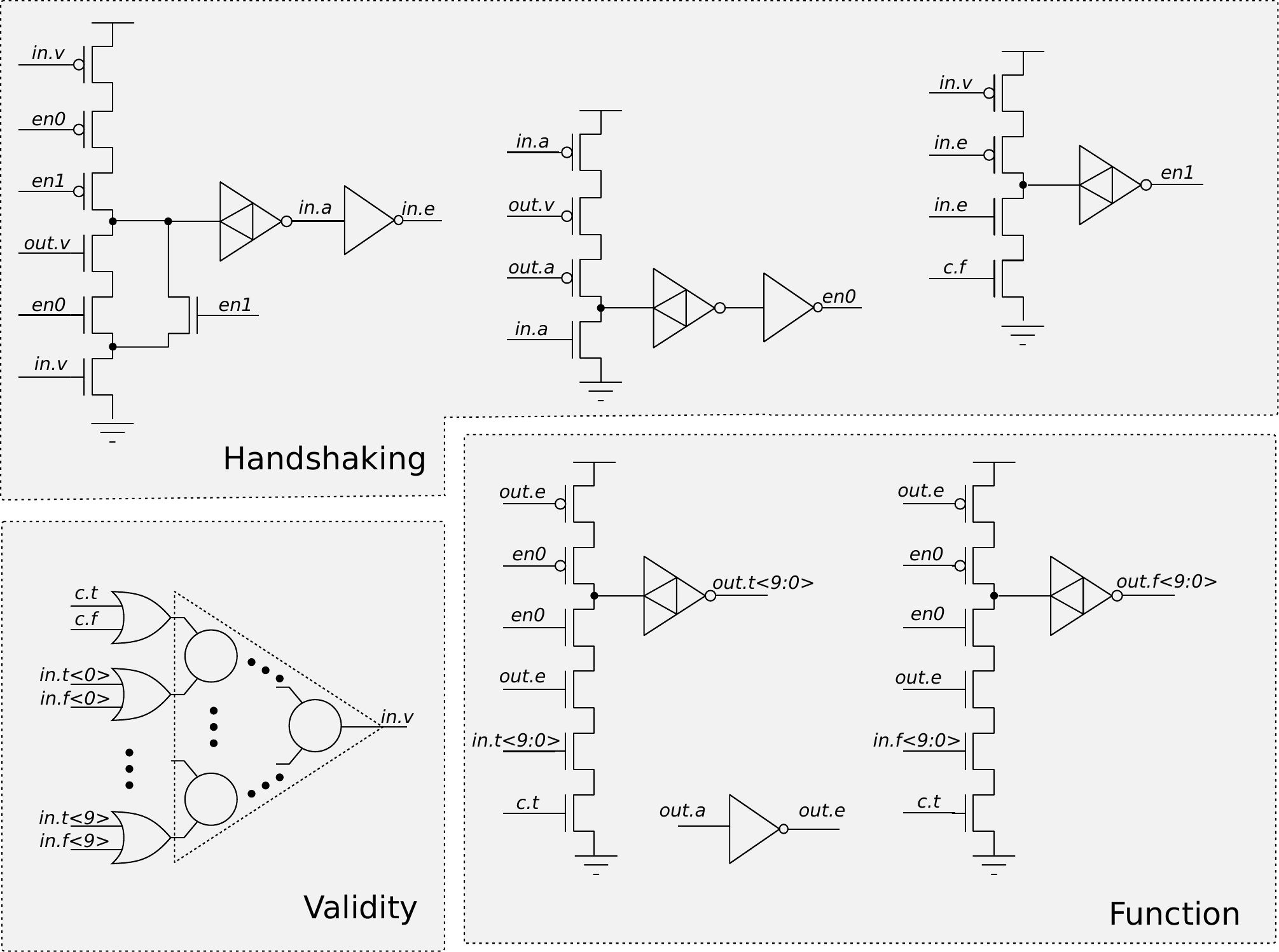}
  \caption{"Controlled-pass" \ac{QDI} circuit diagram. The CMOS circuit is derived based on the production rules of Listing~\ref{list:pr}. \revisednolines{The circles in the circuit represent C-elements.}}
  \label{fig:cn_pass}
\end{figure}

The  asynchronous circuits that implement the R1, R2, and R3 routers, and the overall routing architecture, were synthesized using the \ac{QDI} approach~\cite{Martin90}, by following the \ac{CHP}~\cite{Hoare85} and \ac{HSE} formalism~\cite{Martin_Nystrom06} (see also Appendix~\ref{sec:chp-hse-pr} for the most common \ac{CHP} commands). The \ac{QDI} circuit design methodology only makes timing assumptions on the propagation delay of signals that fan-out to multiple gates, but makes no timing assumption on gate delays.
The \ac{CHP} language provides a set of formal  constructs for  synthesizing a large class of programs~\cite{Manohar06}. Similarly, a Handshaking Expansion program is an intermediate representation of \ac{CHP} constructs in order to close the gap between the high-level representation i.e.\ \ac{CHP} and the low-level circuit description of the system. At this stage of design flow, handshaking protocol and additional restrictions on variable type and assignment statements are introduced into \ac{CHP} program to allow the designer to reach a level of description that is closer to the desired circuit netlist of the system. Finally, \ac{HSE} programs are transformed into a set of ``production rules'' which are abstract descriptions of digital \ac{CMOS} \ac{VLSI} circuits~\cite{Martin_Nystrom06}. As example of asynchronous circuit synthesis, we provide a complete set of \ac{CHP}, \ac{HSE}, and production rule commands for building a \ac{QDI} "Controlled-pass" process. This specific process is commonly used for implementing the routers, and comprises both control flow as well as data-path circuits.
The "Controlled-pass" process has two input channels $in$, $sig$ and one output channel $out$. The $in$ channel holds the input data and the $sig$ channel represents a one bit control signal which decides whether to copy the data from $in$ channel to $out$ channel or not. In other words this process copies the input to the output if $sig$ is \textit{True} and skips otherwise. The $sig$ channel is encoded with dual-rail representation: it uses two physical signals $sig.t$ and $sig.f$ to represent a single bit of information. As the  $sig.t$ and $sig.f$ signals cannot be both \textit{True} at the same time (dual-rail encoding), the decision (pass or not pass) can be made using a \textit{deterministic selection} without the need for arbitration. In the \ac{CHP} program example of Listing~\ref{list:chp}, the process first waits for a new token to arrive on the input channel ($[v(in)]$ becomes true) and checks the value of $sig$ signal. In the case of $sig.t: true$ the process sets the output channel ($out.d\Uparrow$) and waits for the input and the sign signals to become neutral ($[n(in) \land n(sig)]$). Only then the process releases the output signals ($out.d\Downarrow$). If on the other hand $sig.f$ is true, then the process  skips further computations. Figure~\ref{fig:cn_pass} shows the Controlled-pass process circuit diagrams derived from Listing~\ref{list:pr}.

\vspace{1em}
\noindent 
\begin{minipage}{0.475\textwidth}
\begin{lstlisting}[language=chp, caption=controlled pass: CHP, escapeinside={(*}{*)}, mathescape=true, label=list:chp, basicstyle=\footnotesize]
  chp{
    *[[v(in)]; 
    [sig.t -> out.d==; [n(in) ^ n(sig)]; out.d=!;
    []
    sig.f -> skip]; [n(in) ^ n(sig)];
    ]
  }
\end{lstlisting}
\end{minipage}
\hfill
\begin{minipage}{0.475\textwidth}
\begin{lstlisting}[language=chp, caption=controlled pass: HSE, escapeinside={(*}{*)}, mathescape=true, label=list:hse, basicstyle=\footnotesize]
  hse{
    *[[ sig.t -> out.b[i].d[j] := in.b[i].d[j]; in.a+; en0-;
    ([out.a]; out.b[i].d[j]-; [~out.a]), ([n(in)]; in.a-); en0+
    []
    sig.f -> en1+; in.a+; [n(in)]; en1-; in.a-;
    ]]
  }
\end{lstlisting}
\end{minipage}

\begin{lstlisting}[language=chp, caption=controlled pass: PRs, escapeinside={(*}{*)}, mathescape=true, label=list:pr, basicstyle=\footnotesize]
  prs {
    in.v & (en1 | out.v) -> in.a+
    ~en1 & ~en0 & ~in.v & ~sig.f & ~sig.t -> in.a-
    
    ~in.a & ~out.a & ~out.v -> en0+
    in.a -> en0-
    
    in.e & sig.f -> _en1-
    ~in.e & ~in.v -> _en1+
    
    _en -> en1-
    ~_en1-> en1+

    (:i:0..N:
      (:j:0..1:
        en0 & in.b[i].d[j] & sig.t -> out.b[i].d[j]+
        ~en0 & out.e -> out.b[i].d[j]-
      )
    )
  }
\end{lstlisting}

\subsection{Asynchronous routing fabric}
\label{sec:routers}
In the following we describe the block diagrams of the \ac{QDI} routing architecture implemented following \ac{CHP} design methodology. As an example design choice, we assign 256 neurons/core \revised{rev4-cm}{(i.e., $C$ of Fig.~\ref{fig:2-stage} is set to 256)}{}, and four cores/tile \revised{rev4-cm2}{(i.e., $M$ is set to 4)}{}; we assume that each neuron can copy its output to four different destination cores; and, for sake of simplicity, we assume that one tile corresponds to a single \ac{VLSI} chip, which requires only one level of R2 arbiters in the hierarchy. 

\subsubsection{The R1 router}
\label{sec:R1-router}
\begin{figure*}
  \centering
  \includegraphics[width=0.75\textwidth]{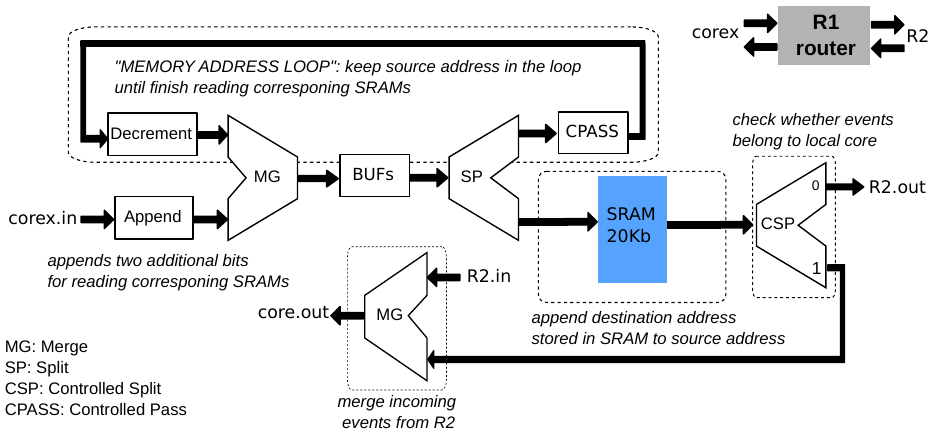}
  \caption{Block diagram of the R1 router. There is one R1 router/core. Each R1 router has a bidirectional link between itself and its local core, and a bidirectional link between itself and its R2 router.}
  \label{fig:R1_arch}
\end{figure*}

Each core in the chip is paired to a R1 router, and each R1 router has an embedded \ac{SRAM} of size proportional to the number of source nodes in the core.
The block-diagram of the R1 router is shown in Fig.~\ref{fig:R1_arch}. Address-events generated by neurons in the corresponding  core ``x'' are sent to R1 (see $corex.in$ signal) The data packet representing the address-event is extended, by appending two additional bits that encode the fan-out for the first-level point-to-point copy (equivalent to $F/M$ in Section~\ref{sec:optimized}).
The extended packet then enters the "memory address loop" of Fig.~\ref{fig:R1_arch}, which comprises a \ac{QDI} merge, a buffer, and a split process, as well as a controlled-pass and a decrement process. The merge process is non-deterministic: it does arbitration on two input requests and allows the winner to send its message through to the output. Initially the input data packet is copied, via the  merge process to a  buffer and then split into two branches: the bottom output of the split process is used to address an asynchronous \ac{SRAM} block, and read the content of the addressed memory; while the upper output of the split process is fed to the controlled-pass process of the feed-back loop. The controlled-pass checks the header bits (the two bits appended in previous stage) and  passes the packet through if the header value is non-zero, otherwise it skips. When passed, the value of header bit is decremented and then merged back to the forward path, in order to read the next address of the \ac{SRAM}. This "memory loop address" process continues until all memory information for the corresponding event is read.

\revised{routing_info}{The content of each \ac{SRAM} cell is a 20-bit word, which includes a 10-bit tag address, a 6-bit header information, and a 4-bit destination core id. The 6-bit header uses 2-bits to encode the $\Delta X$ number of hops, 2-bits for the $\Delta Y$ hops, one sign bit for the $X$ direction~(east or west) and one for the $Y$ directions~(north or south).}{The content of each \ac{SRAM} cell is a 20-bit word, which includes a 10-bit tag address, a 4-bit destination core id (to be used by R2 routers), and a 6-bit chip destination address for long-distance routing (to be used by R3). The 6-bit chip destination address corresponds of 2-bits for the $\Delta X$ jump, 2-bits for the $\Delta Y$ jump, 1-bit for the sign of the $X$ jump (east or west) and 1-bit for the sign of the $Y$ jump (north or south).} A controlled split process after the \ac{SRAM} checks the core and chip destination address bits, and decides whether to broadcast the data back to the same core, to send it point-to-point to a different core on the same chip, or to send it via mesh-routing to a core on a different chip through the R2 and R3 routers. \revised{controlled_split}{The controlled split is a one-input to two-output process where control bits decide which output receives the input data, unlike a normal split process where incoming data are sent to both outputs.}{}

\subsubsection{The R2 router}
\label{sec:R2-router}
\begin{figure*}
  \centering
  \includegraphics[width=0.75\textwidth]{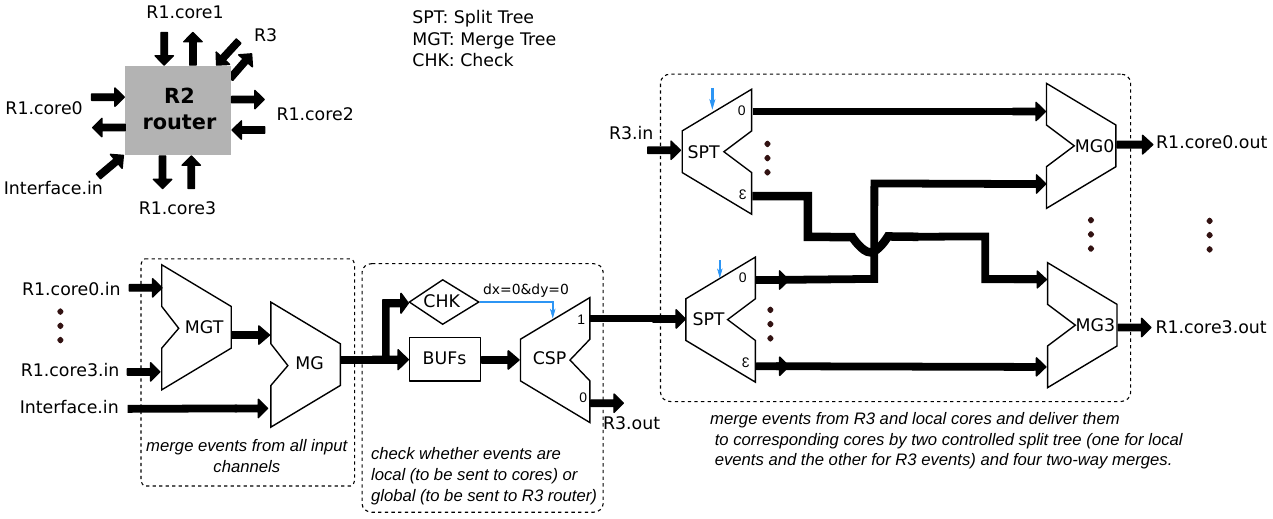}
  \caption{Block diagram of the R2 router. This router has bidirectional links with four R1 routers in the same chip, and one bidirectional link with higher level router for managing inter-chip traffic.}
  \label{fig:R2_arch}
\end{figure*}

Each R2 router has four bidirectional links to R1 routers of the same level of the cores in the chip and one bidirectional link to the higher level router. This router manages both the inter-tile communication and the communication with longer distance targets (see Fig.~\ref{fig:R2_arch}).  Based on a hierarchical 2D tree architecture, this router is at the heart of the proposed routing scheme. At each level of the tree hierarchy there is one bidirectional link between R2 and each core in the same level~(total 4) and one bidirectional link between R2 and the higher-level router. With the design choices we made for these block diagrams, the next level in the hierarchy is directly represented by an R3 router.
Data from all the incoming R1 channels ($R1.core0.in -- R1.core3.in$) is first merged into a single stream by a merge tree. \revised{rev1-2}{Then by}{By then}checking the 6-bit chip destination address entries of the packet, it determines whether to route events to the higher level inter-chip router via $R3.out$ port, or to redirect events to one of the local R1 routers. In the latter case, a controlled split tree decides which local core to target. Similarly, on the downstream pathway, the R2 router receives events form the R3 router (see $R3.in$) and a controlled split tree decides which destination core to target, using the destination core id specified in the packet's header.

\subsubsection{The R3 router}
\label{sec:R3-router}
\begin{figure*}
  \centering
  \includegraphics[width=0.99\textwidth]{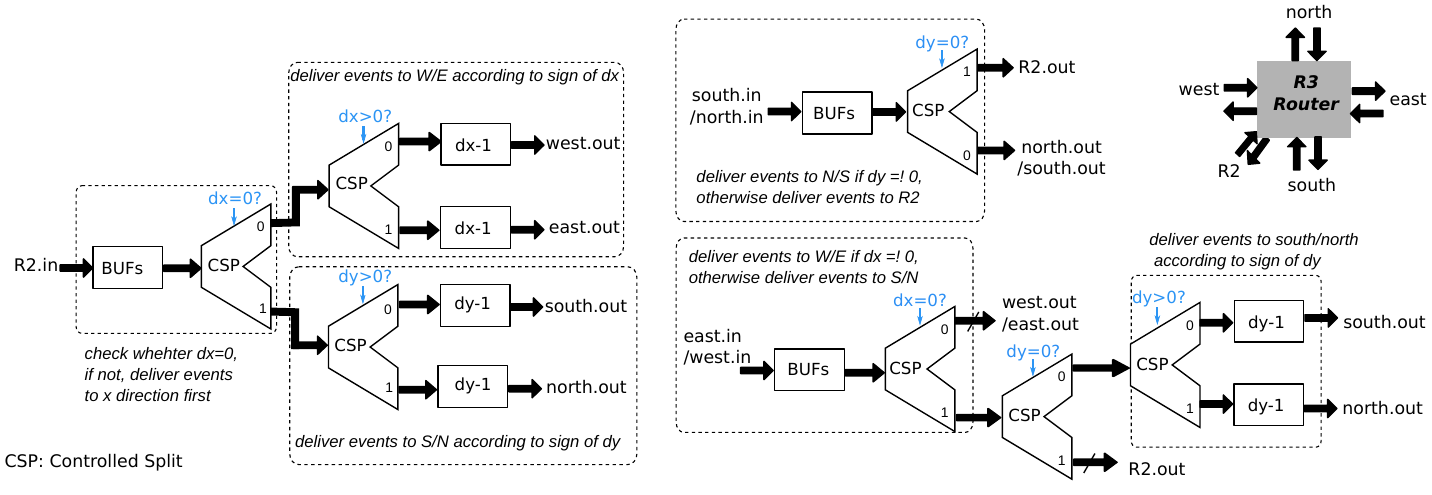}
  \caption{Block diagram of the R3 router. There is one R3 router per chip/tile to manage global traffic in a 2D-mesh of chips/tiles.}
  \label{fig:R3_arch}
\end{figure*}

This block routes events among multiple tiles with relative distance addressing along a 2D-mesh. The block diagram of R3 router is illustrated in Fig.\ref{fig:R3_arch}. On one side, this router communicates with its lower-level R2 router in same tile, and on other side it communicates with other four R3 routers of  other chips/tiles in a global 2D-mesh network \revised{R3-1}{following  an ``XY routing'' method. This method first routes events in X direction until $\Delta X$ is decremented to zero, and then in the Y direction.}{}

\revised{deltaxy}{On the upstream pathway, the router buffers the signals $R2.in$, and performs a control split operation: it first checks  $\Delta X \neq 0$, and if yes sends the data to the west port (if $sign X > 0$) or to the east port (if $sign X < 0$). If the value of $\Delta X = 0$ the events are  sent to the north or south ports, depending on the value of $\Delta Y$. As events are passed through, the corresponding $\Delta X$ and/or $\Delta Y$ values are decremented by one. There are two bits~$sign X$ and $sign Y$ in the packet that specify the sign of $\Delta X$ and $\Delta Y$ changes.}{On the upstream pathway, the router buffers the signals $R2.in$, and performs a control split operation: it first checks if $\Delta X \neq 0$, and if yes sends the data to the west port ($\Delta X > 0$) or to the east port ($\Delta X < 0$). If $\Delta X = 0$ the events are  sent to the north or south ports, depending on the value of $\Delta Y$. As events are passed through, the corresponding $\Delta X$ and/or $\Delta Y$ values are decremented by one.} 
On the downstream pathway, events can arrive from the south/north ports, or from the east/west ones. In the first case, south/north input events trigger a check of the value of $\Delta Y$, done by a controlled split process. If this value is zero, the events are sent to the local R2 router (see $R2.out$ in Fig.~\ref{fig:R3_arch}). Otherwise they are sent to the north/south link. \revised{R3-2}{In the case of incoming events from south/north ports, the $\Delta X$ value is not checked as it will always equal to zero.}{} In the case of an event reaching the east/west port, this triggers a control split process to check the value of $\Delta X$.  If this is not zero, it is passed onto the west/east port and processed as described above, otherwise a check is made on the $\Delta Y$ value, via a second split control process. Depending on the value of $\Delta Y$ the data is passed onto the south/north port, and $\Delta Y$ is decremented.

\subsubsection{Input Interface}
\label{sec:interface}

\begin{figure*}
  \centering
  \includegraphics[width=0.66\textwidth]{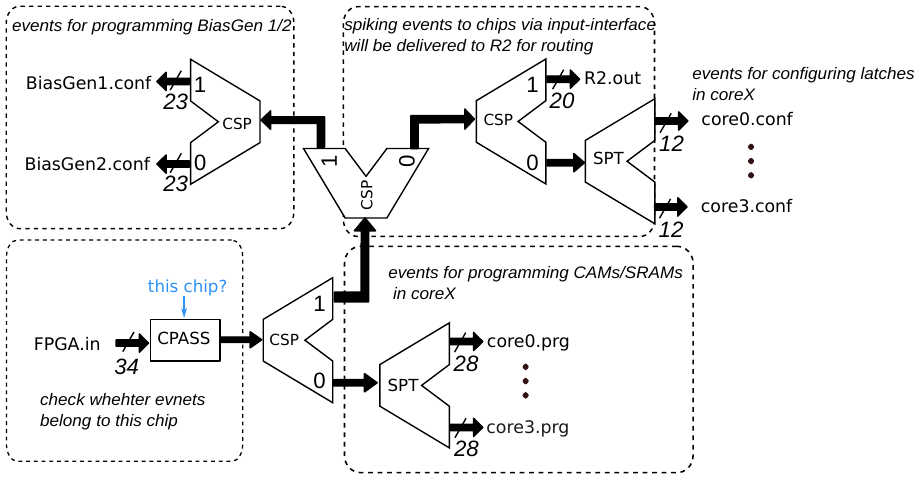}
  \caption{Input Interface block for programming on-chip \ac{SRAM} and \ac{CAM} memories, and for configuring additional network parameters.}
  \label{fig:interface}
\end{figure*}

This additional \ac{QDI} block is used for sending stimuli to the neurons in the cores from external sources of address-events, to program the distributed \ac{SRAM} and \ac{CAM} memory cells,  to set additional network configuration parameters, and to configure different circuit biases (see Fig.~\ref{fig:interface}).
Once address-events reach this block (e.g., from a \ac{FPGA} device, via the $FPGA.in$ signals), a first check is made to see if the data packet contains the correct chip/tile address. If so, the data packet is first checked via a control split process to see if the data represent programming or configuration commands, or normal address-events. In the latter case, the data is sent directly to the destination core. Otherwise the data packet goes through a second control split process to check if it represent control words for programming bias generators, or data for programming the memory cells. Additional control split processes keep on making checks on the data, until it is delivered to its final destination, which could be a register for programming on chip bias generators (for example), or an \ac{SRAM} or \ac{CAM} cell memory content.

\section{A multi-core neuromorphic processor prototype}
\label{sec:hardware}

\begin{figure}
  \centering
  \includegraphics[width=0.48\textwidth]{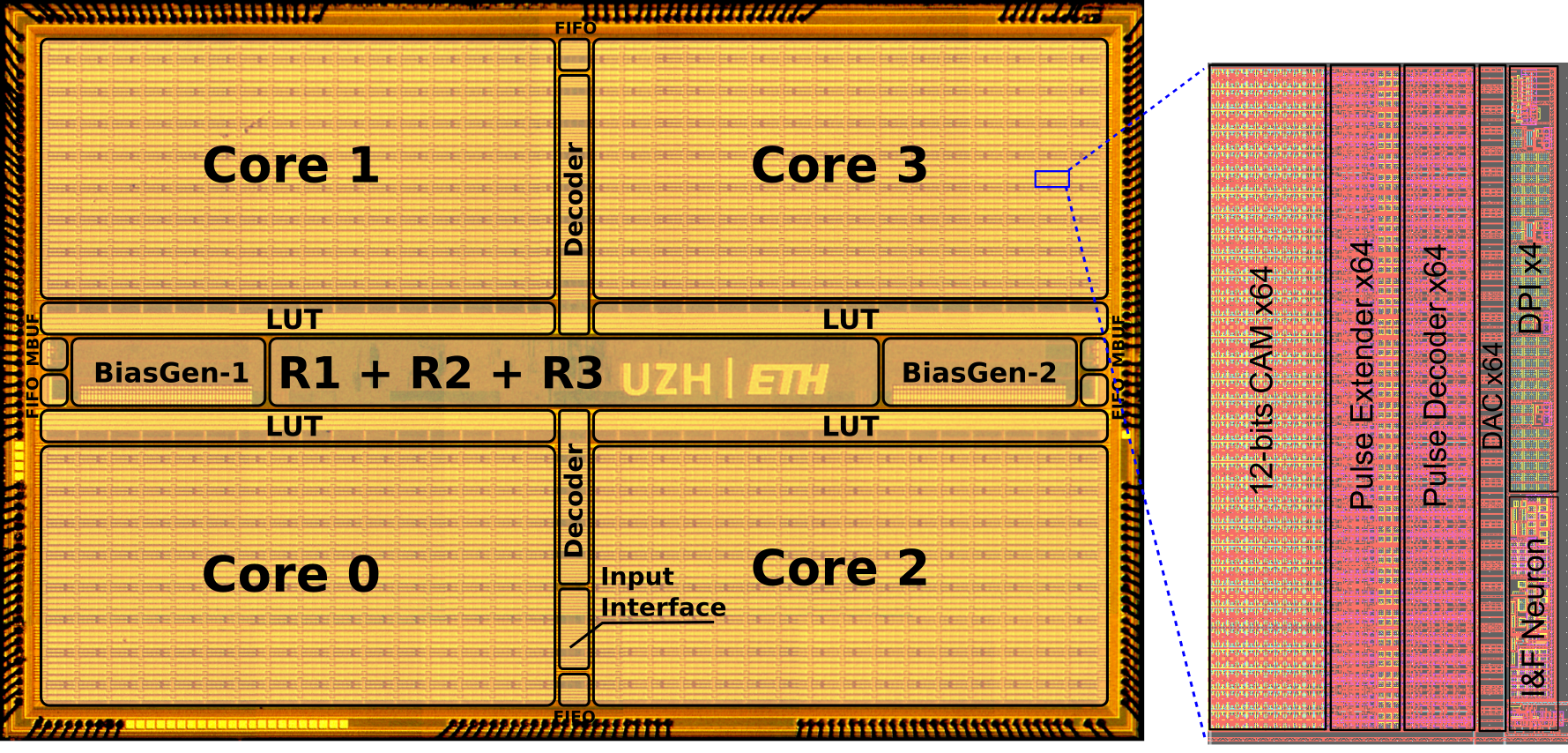}
  \caption{Die photo of the multi-core neuromorphic processor. The chip comprises four cores, each with 256 neurons. Neurons belonging to different cores, and to different chips can interact among each other via the R1, R2, and R3 routers. Neuron and synapse dynamics can be programmed via the on-chip bias generators.}
  \label{fig:die_photo}
\end{figure}

We validated the architecture proposed by designing, fabricating, and testing a prototype  multi-core neuromorphic processor,  made adopting the design choices used in Section~\ref{sec:routers}: the chip comprises four cores; each \revised{fanout}{core comprises 256 neurons, and each neuron has a fan-out of 4k.}{} It has hierarchical asynchronous routers (R1, R2, and R3 level), and embedded asynchronous \ac{SRAM} and \ac{CAM} memory cells, distributed across the cores and the routers. Following the memory-optimizing scheme of Section~\ref{sec:optimized}, the communication circuits combine point-to-point source-address routing with multi-cast destination-address routing, and use a mixed-mode hierarchical-mesh connectivity fabric.

The chip was fabricated using a standard 0.18\,um 1P6M \ac{CMOS} technology, and occupies an area of 43.79\,mm$^2$ (see Fig.~\ref{fig:die_photo} for the chip micro-graph).  The memory optimization theory and the hierarchical routing fabric can be used with either pure digital logic approaches~\cite{Merolla_etal14a}, or mixed mode analog/digital ones~\cite{Chicca_etal14}. Here we demonstrate a neuromorphic processor designed following a mixed signal approach: we implemented the neuron and synapse functions and state-holding properties using parallel analog circuits, rather than time-multiplexed digital ones. The analog circuits are operated in the sub-threshold domain to minimize the dynamic power consumption and to implement biophysically realistic neural and synaptic behaviors, with biologically plausible temporal dynamics~\cite{Chicca_etal14}.

\begin{table}
  \centering
  \begin{tabular}{l  l}
    \toprule
    \textbf{Block name} & \textbf{Silicon area\%}\\
    \midrule
    On-chip memory & 31.7\,\%\\
    Pulse extenders and decoders & 25.2\,\%\\
    Neurons and synapses & 22.8\,\%\\
    Routers & 9.7\,\%\\
    Bias generators & 6.4\,\%\\
    Other & 4.3\,\%\\
    \bottomrule
  \end{tabular}
  \caption{Detailed area breakdown of the chip's circuit blocks. The core area of the chip is 38.5\,$mm^2$.}
  \label{tab:area}
\end{table}

The core area of the chip layout (excluding pad frame) measures 38.5\,$mm^2$,  of which approximately 30\% is used for the memory circuits, and 20\% for the neuron and synapse circuits (see Table.~\ref{tab:area} for a detailed breakdown of the area usage). Neurons are implemented using \ac{ADEX} neuron circuits of the type described and fully characterized in~\cite{Qiao_etal15}. The circuit comprises a block implementing \ac{NMDA} like voltage-gating, a leak block implementing neuron’s leak conductance, a negative feedback spike-frequency adaptation block, a positive feedback block which models the effect of Sodium activation and inactivation channels for spike generation, and a negative feedback block that reproduces the effect of Potassium channels to reset the neuron's activation and implement a refractory period.
The negative feedback mechanism of the adaptation block and the tunable reset potential of the Potassium block introduce two extra variables in the dynamic equation of the neuron that endow it with a wide variety of dynamical behaviors~\cite{Izhikevich06,Naud_etal08}.

Synapses and biophysically realistic synapse dynamic are implemented using sub-threshold \ac{DPI} log-domain filters, of the type proposed in~\cite{Bartolozzi_Indiveri07a}, and described in~\cite{Chicca_etal14}. These circuit can produce \acp{EPSC} and \acp{IPSC} with time constants that can range from few $\mu$s to hundreds of ms, using relatively small capacitors (of the \revised{}{value}{size} of about 1\,pF), thus keeping the circuit footprint to a minimum size.

The analog circuit parameters governing the behavior and dynamics of the neurons and synapses are set by programmable on-chip temperature compensated bias-generators~\cite{Delbruck_etal10}. There are two independent bias-generator blocks, to provide independent sets of biases for core pairs, to allow the modeling of different neural population types.
The use of mixed-mode analog/digital circuits, embedded in the asynchronous routing fabric proposed  allowed us to distribute the memory elements across and within the computing modules, creating an in-memory computing architecture that makes use of distributed and heterogeneous memory elements (such as capacitors, \ac{CAM}, and \ac{SRAM} cells) and that is truly non-von Neumann. Alternative neuromorphic processors that use pure digital design styles typically time-multiplex the computing resources and are still faced with the problem of having to transfer state memory back and forth from areas devoted to computing to areas devoted to memory storage~\cite{Merolla_etal14a,Furber_etal14}.

\subsection{The core memory/computing module}
\label{sec:node}

\begin{figure*}
  \centering
  \includegraphics[width=0.78\textwidth]{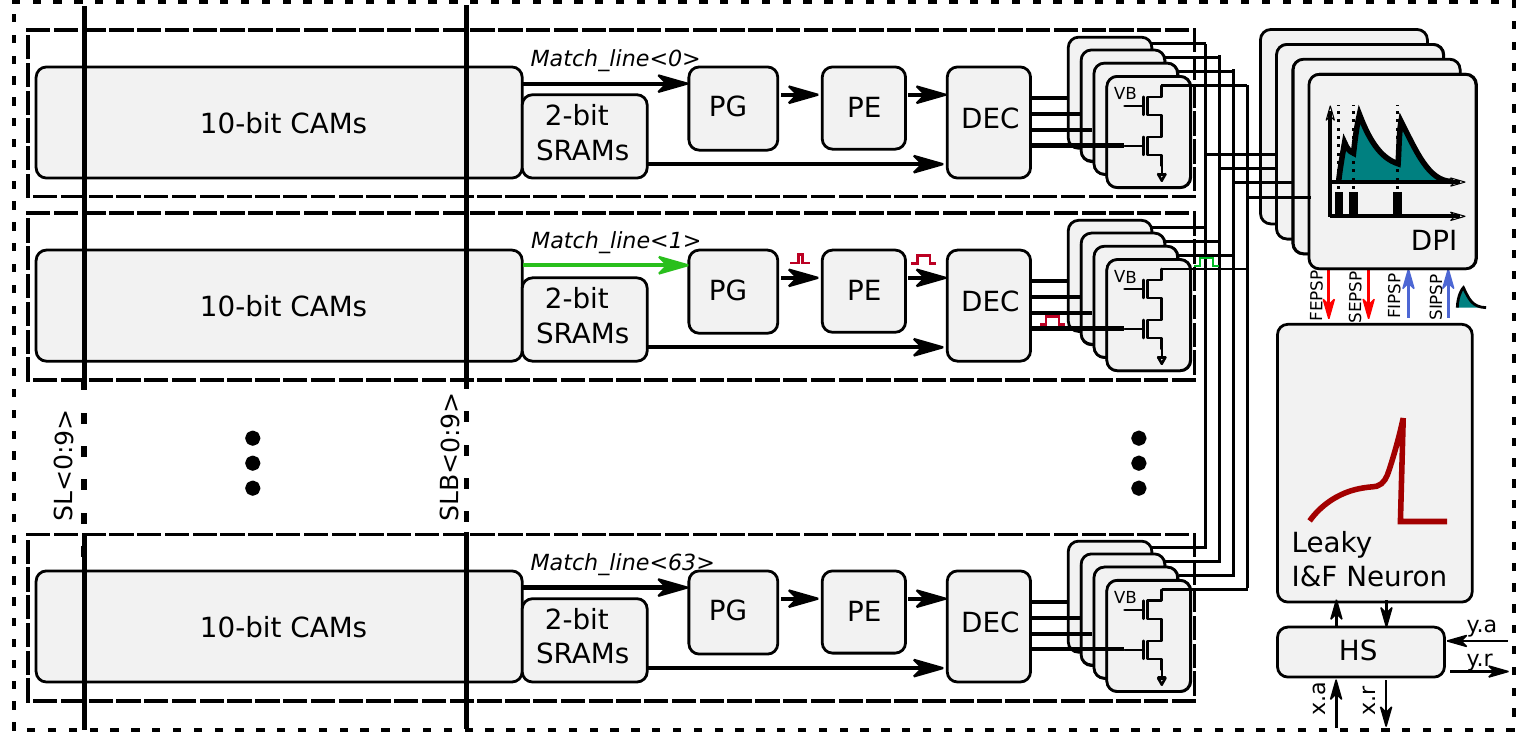}
  \caption{Block diagram of one of the 256 computing nodes in each core. Each node comprises 64 mixed memory words, consisting of 10-bit \ac{CAM} and 2-bit \ac{SRAM} cells, 64 pulse generators circuits (PG), 64 pulse extenders circuits (PE), 64 pulse decoders (DECs), \revisednolines{64$\times$4 digital pulse to analog current converters, 4 \ac{DPI} filters}, one adaptive I\&F neuron circuit, and one handshaking block (HS).}
  \label{fig:computing_node}
\end{figure*}
The block diagram of the circuits comprising the synapse memory and synapse dynamic circuits together with the neuron circuits is shown in Fig.~\ref{fig:computing_node}. Each of these nodes implements at the same time the memory and computing operations of the architecture: \revised{rev4-38}{there are 64 10-bit \ac{CAM} words, 64 2-bit \ac{SRAM} cells, ``four'' synapse circuits, and ''one'' leaky integrate and fire neuron circuit per node.}{each neuron has synapse circuits connected to 64 10-bit \ac{CAM} cells and 64 2-bit \ac{SRAM} cells.} The asynchronous \ac{CAM} memory is used to store the tag of the source address that the neuron is connected to, while the 2-bit \ac{SRAM} memories are used to program the type of synapse circuits to use.  Depending on the content of the \ac{SRAM} a synapse can be programmed to exhibit one of 4 possible behaviors: fast excitatory, slow excitatory, subtractive inhibitory, or shunting inhibitory. Each synapse behavior is modeled by a dedicated \ac{DPI} circuit, each with globally shared biases that set time constants and weight values.
When a synapse accepts an address-event that has been broadcast to the core (i.e., when there is a match between the address transmitted and the one stored in the synapse \ac{CAM} cell),  the event triggers a local pulse-generator circuit, which in turn drives a pulse-extender circuit to produce a square wave of tunable width, ranging from fractions of $\mu$s to tens of ms. These square waves are then fed into the addressed \ac{DPI} circuit, which integrates them over time, and produces an \ac{EPSC} or \ac{IPSC} with corresponding weight and temporal dynamics.

Eventually, once all synaptic currents integrated by the neuron make it cross its spiking threshold, the neuron will produce an output digital event and send it to its handshaking block (HS). This block will communicate with neuron address encoder which will transmit the address-event to the core associated R1 router.



\subsection{Asynchronous Content-Addressable Memory circuits}
\label{sec:cam}
\begin{figure*}
\centering
\includegraphics[width=0.95\textwidth]{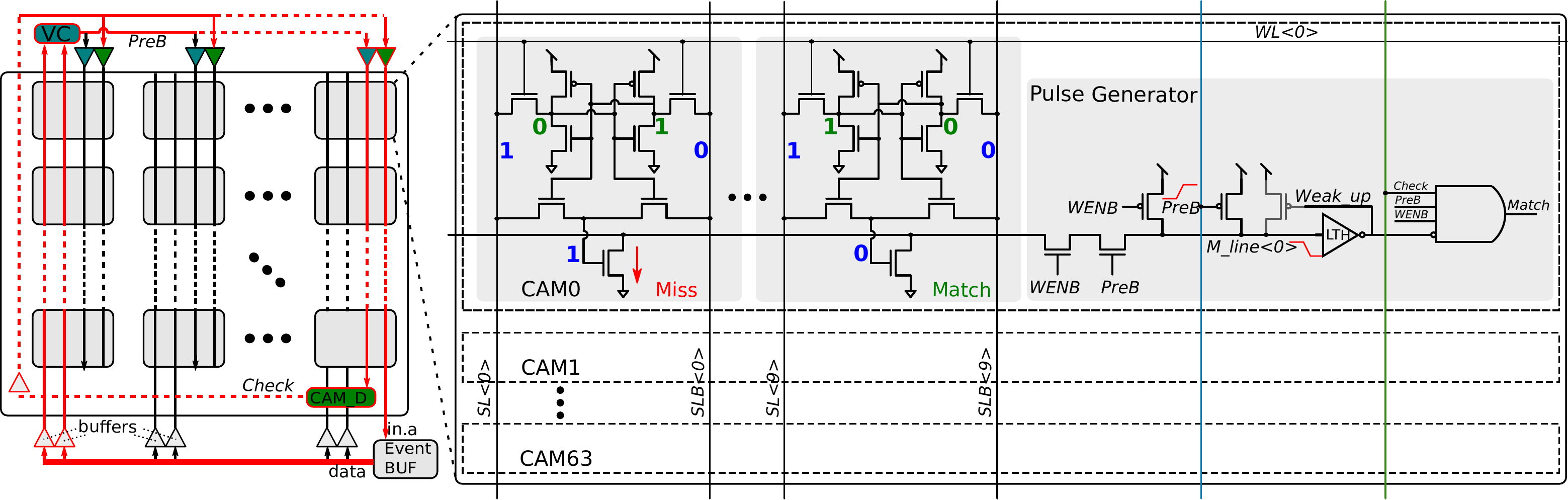}
\caption{
Asynchronous Content-Addressable Memory (CAM) circuit diagram.
Left: a simplified $16\times16$ \ac{CAM} blocks in one core each having 64 words corresponding to 64 active synapses of a neuron.
Right: Circuit details of one \ac{CAM} word combined with one pulse generator circuit.
}
\label{fig:cam}
\end{figure*}

Figure~\ref{fig:cam} shows a simplified diagram of the $16\times16$ asynchronous \ac{CAM} cells, as they are arranged in each core. Each \ac{CAM} cell comprises 64 10\,bit word \ac{CAM} circuits, as described on the left side of Fig.~\ref{fig:cam}, which contain the addresses of 64 source neurons connected to the corresponding destination neuron's synapses.
The \ac{CAM} cells make use of NOR-type 9T circuits and of a pre-charge-high Match-Line scheme~\cite{Pagiamtzis_Sheikholeslami06}.
In the pre-charge phase, no data is presented on the search bit lines (neutral state of data) and the signal $PreB$ is asserted to {\em low}, in order to pre-charge all the Match-Lines in the core to {\em high}. 
In the search phase, $PreB$ is asserted to {\em high} and all \acp{CAM} compare their contents with the data presented on the search lines simultaneously.
A {\em miss} between any input data bit and the content of the corresponding \ac{CAM} bit circuit discharges the corresponding match line.
Thus, only \ac{CAM} words with a {\em match} on all bits keep the corresponding Match-Line {\em high}.
A Low-Threshold  (LTH) inverter together with a weak pull-up P-MOS transistor is used to compensate the leak from the {\em NOR} transistors during the search phase in order to guarantee that only a real {\em miss} will pull the  Match-Line low.
After all \ac{CAM} comparisons are complete, a pulse signal $Check$ is transmitted across the whole core and multiplied with the Match-Lines (i.e., via a logical {\em AND} operation)  to generate pulses for all \ac{CAM} words that are in a {\em match} state.
The matched pulses are then transmitted to the pulse generation circuit of Fig.~\ref{fig:computing_node}, and used to produce the synaptic currents.

The search operation of the asynchronous \ac{CAM} array follows a standard four-phase handshaking protocol, to communicate with the \ac{QDI} routing blocks. However, to minimize its area, the interface to the \ac{CAM} array make timing assumptions and does not implement a \ac{QDI} design style. The full description of the asynchronous \ac{CAM} circuit behavior, and of the timing assumptions made are detailed in Appendix~\ref{sec:timing-assumptions}


\section{Experimental results}
\label{sec:experimental-results}

The 0.18\,um \ac{CMOS} process used to fabricate the chip expects a core power-supply voltage of 1.8V, however thanks to a careful design of the analog and asynchronous digital blocks, we could reduce the core supply voltage down to 1.3V without any loss of functionality.
The chip specifications, including speed and latency measurements, are outlined in Table~\ref{tab:features}. As evident from this table, although the asynchronous routers deliver a high throughput within the chip, the overall system performance is limited by the speed of the I/O circuits. In larger designs this would not be a major problem, as the communication among neurons would mostly happen inside the chip. The other restricting factor is the broadcast time, i.e., the time that the \ac{CAM} requires to process the incoming events. Also in this case, and similar to  most memories designs that make use of asynchronous interfaces, we made worst-case scenario timing assumptions that guarantee correct functionality of the \ac{CAM} circuits block~(see Appendix for details). In the current design, we set the broadcast time to be 27\,ns. Also this will not be a critical limitation in large scale multi-core designs, as the \ac{CAM} cells of different cores operate in parallel. 

\begin{table}
  \renewcommand{\arraystretch}{1.3} 
  \centering
  \begin{tabular}{l  l}
    \toprule
    Process Technology & 0.18\,um 1P6M\\ 
    Supply voltage (core) & 1.3\,V-1.8\,V\\
    Supply voltage (I/O) & 1.8\,V-3.3\,V\\
    Die Size & 43.79\,mm$^2$ \\ 
    Number of Neurons & 1\,k\\   
    Number of Synapses & 64\,k\\ 
    Total Memory & 64\,k \acs{CAM} + 4\,k \acs{SRAM}\\ 
    I/O Speed & 30\,M\,events/s (input) / 21\,M\,events/s (output)  \\
    LUT Read Speed  & 750\,Mb/s\\ 
    Broadcast time$^1$   & 27\,ns\\ 
    Latency across chips$^2$ & 15.4\,ns\\
    \bottomrule
  \end{tabular} 
  \caption{Basic specifications of the multi-core neuromorphic processor, $^1$ The broadcast time refers to the time it take to broadcast an incoming event to the core. It includes the broadcast buffer, the \ac{CAM} search and the handshaking times. $^2$ The latency for passing through the chip. It involves the R3 router and on-chip interconnects latency, taking in account the capacitance load of the output channel.}
  \label{tab:features} 
\end{table}

\begin{figure}
\centering
\includegraphics[width=0.44\textwidth]{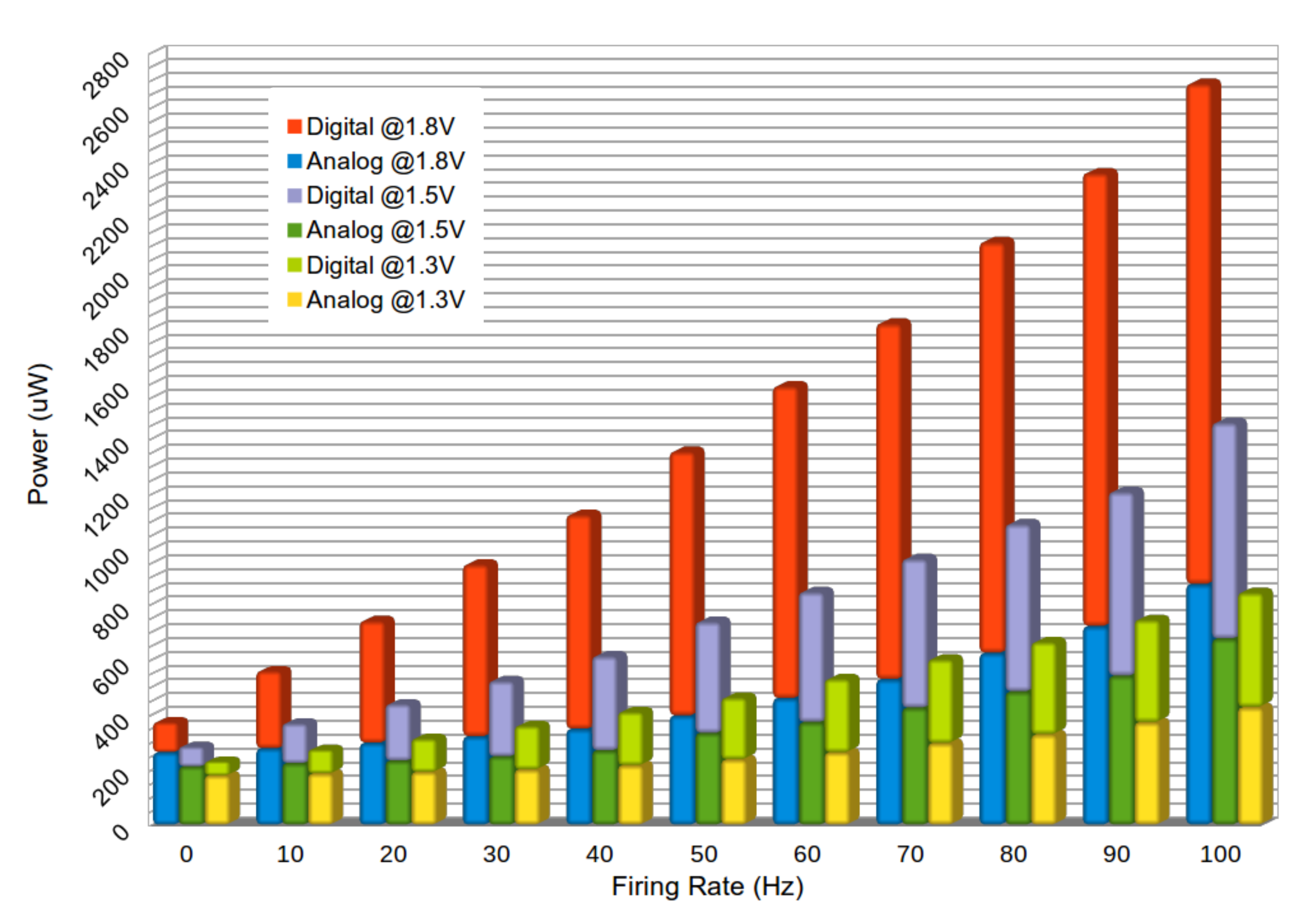}
\caption{Power dissipation measured for different supply voltage settings as a function of different average firing rates of all the neurons of the chip (worst case scenario). The firing rate is set by injecting a constant DC current to the neuron membrane capacitance. In this experiment, all neurons have active connections to 25\%  of all neurons in the chip, assuming the destination neurons for each spike resides across four cores. This measurement does not include the effect of synaptic currents (synaptic input currents neurons are replaced by the DC current injection for better control of their average firing rate\label{fig:power_dissipation}}
\end{figure}

Figure~\ref{fig:power_dissipation} shows the average power dissipation measures, for different power-supply voltage settings, in a worst-case scenario condition in which all neurons of the chip are firing, at different average firing rates. The power usage of the major circuits in the neuromorphic processor is reported in Table~\ref{tab:power}. 

\begin{table}[hb]
  \centering
  \begin{tabular}{l c c}  
    \toprule
    \textbf{Operation} & 
    \textbf{1.8\,V} & 
    \textbf{1.3\,V} \\ 
    \midrule
    Generate one spike & 883\,pJ & 260\,pJ\\ 
    Encode one spike and append destinations & 883\,pJ & 507\,pJ\\
    Broadcast event to same core & 6.84\,nJ & 2.2\,nJ\\
    Route event to different core & 360\,pJ & 78\,pJ\\
    Extend pulse generated from \ac{CAM} match & 324\,pJ & 26\,pJ\\    
    \bottomrule
  \end{tabular} 
  \caption[]{Energy Consumption of the main operations}  
  \label{tab:power} 
\end{table}

Despite the use of a 0.18\,um \ac{CMOS} process, and thanks to the mixed signal analog/digital and asynchronous design strategies adopted, the chip  proposed achieves power consumption and latency figures that are comparable to analogous state-of-the-art architectures fabricated with more advanced scaled technology nodes. In Table~\ref{tab:measurement} we provide a detailed comparison between the specifications of this neuromorphic prototype chip and of recently proposed analogous neuromorphic systems.

\begin{table*}
  \centering
  \begin{tabular}{l c c c c c}  
    \toprule
    \textbf{} & 
    \textbf{IBM Truenorth~\cite{Merolla_etal14a}} & 
    \textbf{Spinnaker~\cite{Furber_etal14}} & 
    \textbf{HiAER~\cite{Park_etal16}} & 
    \textbf{Neurogrid~\cite{Benjamin_etal14}} & 
    \textbf{This work--DYNAPs}\\ 
    \midrule
    Technology & 28\,nm  & 0.13\,um & 0.13\,um & 0.18\,um & 0.18\,um\\ 
    Neuron Type & Digital & Digital & Mixed & Mixed & Mixed\\
    In core computation & Time multiplexing & Time multiplexing & Parallel & Parallel & Parallel \\
    Fan-in/out & 256/256 & /1k & x/1k & x & 64/4k\\
    Routing & on-chip & on-chip & on/off chip & on/off chip & on-chip\\
    Energy per hop & 2.3\,pJ@0.77\,V  & 1.11\,nJ & x& 14\,nJ & 17\,pJ@1.3\,V\\
     Avg. Distance* & $2\sqrt{N}/3$  & $\sqrt{N}/2$& x& x & $\sqrt{N}/3$\\ 
    \bottomrule
  \end{tabular} 
  \caption{Features of the multi-core neuromorphic processor presented in this work, and other recent analogous devices. *Average distance between any two nodes in the network of $N$ neurons; HiAER and Neurogrid, hierarchical and tree-based network respectively, have utilized a combination of on-chip and off-chip resources for routing, therefore it is unclear what is the average distance between the nodes. }  
  \label{tab:measurement} 
\end{table*}
\revised{extra}{
The throughput of the local event-delivery circuitry and the routing networks are important factors for the overall scalability of the proposed architecture. At the on-chip core level, the throughput is determined by the broadcast time. In the current implementation the broadcast time is of $\approx$27\,ns (leading to a bandwidth of $\approx$38\,Mevents/sec) and the number of neurons per core i.e. 256. This results in a throughput that allows us to have 7200 fan-in per neuron in a network with the average firing rate of 20\,Hz and 1400 fan-in at 100\,Hz. As these are digital circuits, the throughput in the local core is expected to  improve significantly when implemented in more advanced processes. 
At the large-scale network level, the ``latency across chip'' figure~(which measures $\approx$15.4\,ns in the current implementation) determines how many cores can be reliably integrated in a \ac{PCB} board. As  this number is experimentally measured, it includes the latency of  input pins, R3 router and  output pins. In a many-core large chip design, the latency of the I/O pins would no longer impact the R3 throughput. In our 0.18\,$\mu$m prototype, R3 has the latency of 2.5\,ns, delivering 400\,Mevent/sec. According to circuit simulation studies~\cite{Qiao_Indiveri16}, the throughput of the R3 router can reach up to 1\,Gevents/sec in a 28\,nm process.      
Thanks to the hierarchical structure of the routing architecture, a high percentage of local activity of a clustered network is sorted out by the R2 and R1 routers without need to involving the R3 routers. Therefore the traffic at the top level of hierarchy significantly decreases in comparison to that of a plain 2D mesh architecture. Furthermore, being implemented using only split/merge and decrements blocks, without any look-up table or feedback mechanism, the R3 router was carefully designed to have a simple and feed-forward structure. This structure provides more opportunity to increase the throughout by increasing the number of pipeline stages in future revision of this work.}{}
\subsection*{Example application: Convolutional Neural Networks}
\label{sec:cnn}
The architecture we proposed, and the prototype chip implementation used to validate it can be applied to a wide range of network types, ranging from biophysically realistic models of cortical networks, to more conventional machine learning and classical artificial neural networks. Here we use the prototype chip designed to implement \acp{CNN}. \acp{CNN} are a class of models originally inspired by the visual system, that are typically used to extract information from input images for classification.
They are multi-layered networks whereby low-level features, such as edge orientations extracted by early layers, are subsequently combined to produce high-level feature detectors in later stages. \revised{board}{To support experiments with multi-layer network structures, we developed a \ac{PCB} board which hosts nine chips (of the type shown in Fig.~\ref{fig:die_photo}). The inter-chip communication on the board is carried out by parallel input/output ports through direct wire/pin connections. The communication is directly managed by the programmable on-chip routers~(R3). An additional on-board \acs{FPGA} device is used for programming the on-chip memories, configuring the analog parameters and monitoring the neuron activities from all chips. The board is extendable as it has connectors on the four sides for interfacing to other instances of the same \ac{PCB}.}{}
This feature extracting process makes the classification problem of the later stages much simpler. Therefore neural networks that combine several hidden layers of this form, known as deep neural networks (DNNs), can be efficiently trained to achieve very high performance on a wide range of visual and non-visual tasks~\cite{LeCun_etal15}.
Recently, pre-trained CNNs and DNNs have been successfully mapped into spike-based hardware, achieving remarkable accuracy, while minimizing power consumption~\cite{Diehl_etal15, Cao_etal15, Marti_etal16, Esser_etal16}. Furthermore, there are growing efforts towards the direct use of asynchronous event-based data produced by spiking sensors such as the \ac{DVS}~\cite{Perez-Carrasco_etal13, Serrano-Gotarredona_Linares-Barranco15, Henderson_etal15, Zhao_etal15, Orchard_etal15, Lagorce_etal16}.
Here we tailored the design of the \ac{CNN} for efficient real-time classification of event streams using an \ac{AER} data-set recently made available~\cite{Serrano-Gotarredona_Linares-Barranco15}.
This data-set consists of a list of time-stamped events generated by an \ac{AER} dynamic vision sensor while observing a very fast sequence of Poker cards flipped from a deck. 
All 52 cards of the deck are flipped in front of the vision sensor in about half a second producing a total of 0.5 million events with a peak rate of slightly above 8 Meps~\cite{Camunas-Mesa_etal12}. The events obtained in this way have been further processed to center the Poker card symbol in the center of 31$\times$31 image patches.
\begin{figure*}
    \centering
    \includegraphics[width=0.78\textwidth]{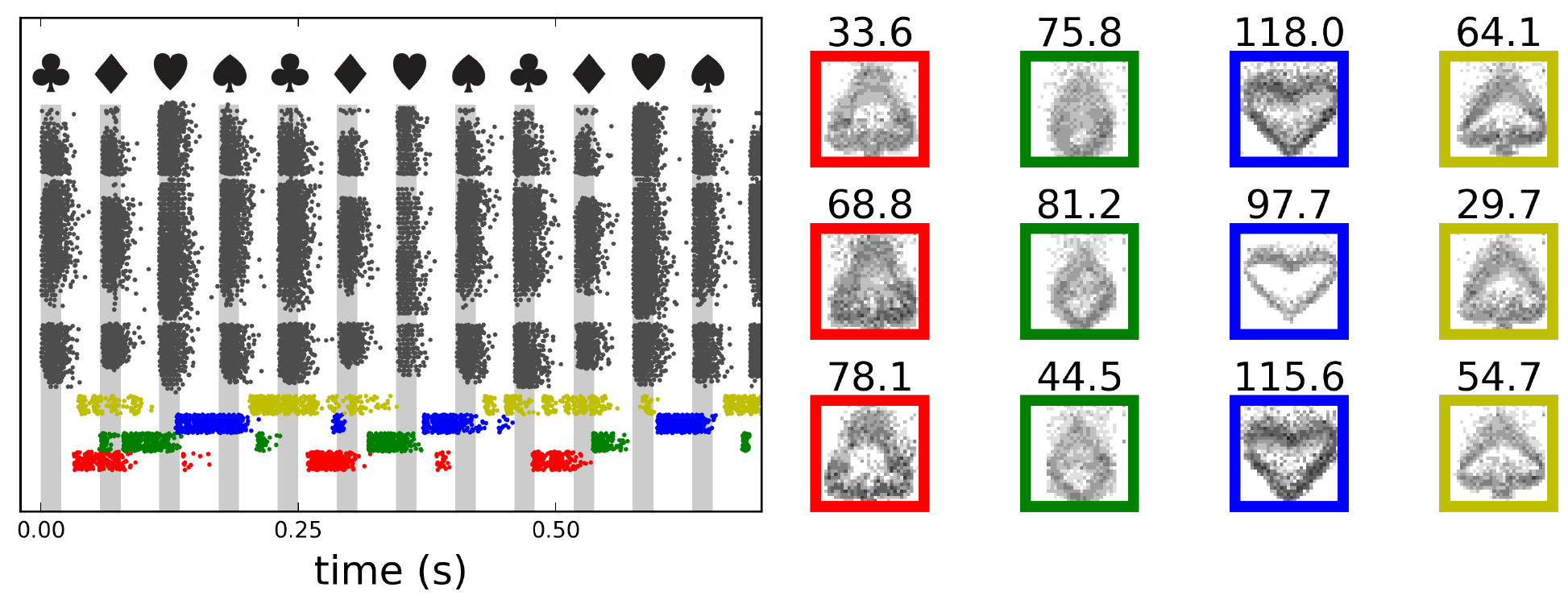}
  \caption{\ac{CNN} results of Poker card suit recognition experiment.
Left panel: raster plot of input (grey) and output (colored) units.
The symbols corresponding to the presented patterns are shown on the top.
Right panel: image frames computed by aggregating the spiking outputs over time.
In each panel, spikes aggregated from the input are shown  with spike frequency encoded in grey levels, from 0 (white) to 400 Hz (black).
The spiking data for each panel corresponds to the vertical grey bars in the raster plot on the left.
The color of the panel borders encodes the estimated Poker card suit. It is determined by identifying the most active population of neurons in the output layer (with corresponding firing rate in Hz on top of each panel). The firing rates reported have been determined by the spike count in a \revisednolines{20\,ms}{} window delayed by \revisednolines{24\,ms}{} from the time window of the input to consider the delaying effects of synaptic integration.
The input streams for each symbol have been manually separated for visual clarity.}
\label{fig:experiment}
\end{figure*}
Their weights have been set to detect vertical and horizontal edges as well as upward and downward vertices.
The resulting 16x16x4 maps are then pooled into a layer comprising 4x8x8 maps.
The activity of the silicon neurons in the pooling layer is then used as input to four populations of 64 neurons in the output layer.
The all-to-all connections between pooling layer and output layer are tuned using an off-line Hebbian-like learning rule such that, \revised{rev1-3}{for}{foe} each input symbol, the 64 most active pooling layer neurons are strongly connected with the corresponding output neurons subgroup.
We used 4 populations of 64 neurons, rather than just 4  neurons in the output layer,  to stabilize the performance, by using a majority rule. The  class corresponding to the right Poker card suit is thus determined by the most active population, during the  presentation of the input pattern (see Fig.~\ref{fig:experiment}).
As evident from Fig.~\ref{fig:experiment}, the asynchronous event-based  processing nature of this architecture allowed the system to produce output results with extreme low-latency figures: the classification result is available within less than 30~ms from the onset of the  input stimulus. This figure was obtained without optimizing the network parameters for speed \revised{rev1-4}{.}{} The time constants in the analog circuits can be tuned to be much faster than the biologically plausible ones used in this demonstration (i.e., in the order of tens of milliseconds).
\begin{table}[hb]
  \centering
  \begin{tabular}{l c c}  
    \toprule
    \textbf{Parameter} & \textbf{Size} \\ 
    \midrule
    Input image & 32$\times$32 \\ 
    Conv. kernels & 4$\times$8$\times$8, Stride=2 \\
    Conv. layer output& 4$\times$16$\times$16 \\ 
    Subsampling kernels & 2$\times$2\\
    Subsampling output & 4$\times$8$\times$8 \\
    Fully-connected layer& 4$\times$64 \\
    \bottomrule
  \end{tabular} 
  \caption[]{An example CNN for Poker card suit recognition experiment.}  
  \label{tab:CNNarch} 
\end{table}
The simplicity of the chosen problem was such that even a simple architecture such as the one described here, that used merely 2560 neurons was sufficient to achieve a 100\% performance on the test data set. Multi-chip boards, such as the 9-chip board used for this experiment, can support much larger and sophisticated networks (of up to 9k neurons). Indeed, it has been recently shown~\cite{Diehl_etal15,Marti_etal16} that if these networks are designed appropriately, they can tolerate the limited  precision that affects the analog circuits in the chip, while preserving high accuracy figures that are comparable to those obtained by state-of-the-art \acp{DNN} running on hardware consuming orders of magnitude more power.

\begin{figure}[htbp]
  \centering
  \includegraphics[width=0.45\textwidth]{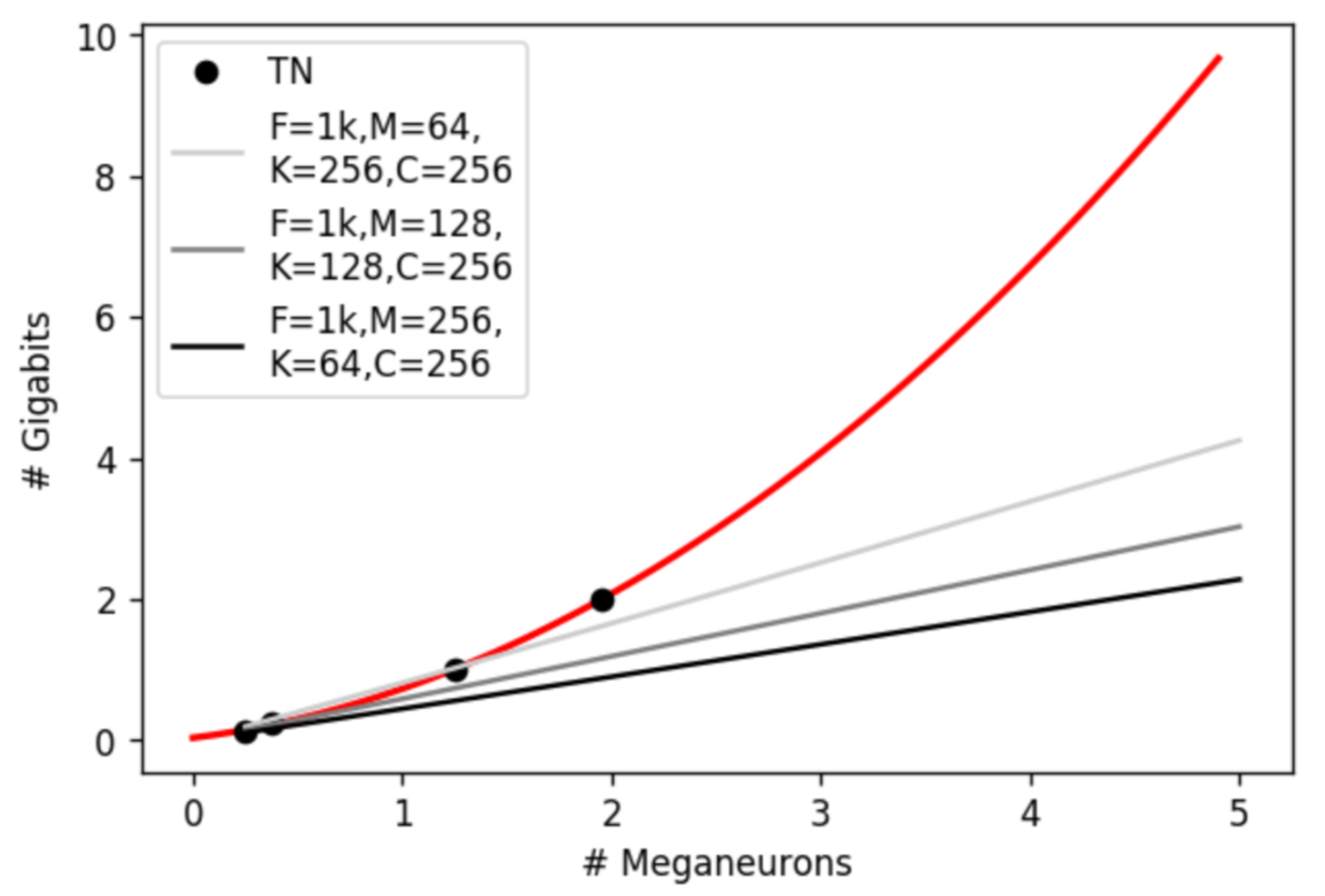}
  \caption{\revisednolines{Memory scaling comparison between the TrueNorth architecture~\cite{Merolla_etal14a} and this work. The number of bits used by the two architectures is plotted as a function of the \ac{CNN} model size. For TrueNorth, 4 data points (black dots) are extrapolated from CNN benchmark models described in~\cite{Esser_etal16} and  fitted with a quadratic function (red curve). The number of cores, and therefore the number of bits, used to implement the 4 models scales approximately quadratically with the size of the models (red line). This is likely due to the fact that in this architecture extra neuro-synaptic cores need to be  used to expand neuron fan-in and fan-out for routing. In the proposed architecture instead, no additional cores are required for routing. Hence the scaling is linear with model size (dark lines, different shades represent different parameter choices). 
      The prototype chip fabricated uses $KM/C=64$. The scaling plots were computed from eq.~(\ref{eq:bits-neuron}), but adding 2 extra bits per neuron for 4 synaptic weight types as in~\cite{Esser_etal16}.}}%
  \label{fig:tn_scaling}
\end{figure}

One important and novel aspect of the hardware and framework proposed in this work is the flexibility offered by the  memory optimized programmable routing scheme, that allows an efficient allocation of memory, synapse, and neural resources for a wide range of architectures, including \acp{CNN}. This is evident if one analyzes the amount of resources required by comparable architectures, such as the one proposed in~\cite{Merolla_etal14a}, for different types of \acp{CNN} architectures, as recently reported in~\cite{Esser_etal16}.
We analyzed the memory size for CNN models implemented on TrueNorth.
The type of neuron and synapse models implemented in the two hardware platforms are very similar, therefore we'd expect similar classification performance if those networks were to be implemented on the proposed platform.
Here we consider the scaling that we would obtain if we were to implement those CNN networks on our hardware platform.
For TrueNorth, we observed a roughly quadratic relation between the number of hardware cores used and the number of neurons required by the network, despite the filter size being fixed by design in agreement with the hardware constraints (Fig.~\ref{fig:tn_scaling}).
This scaling comes from the fact that in TrueNorth additional cores are allocated for implementing larger fan-out as required by the models.
Instead, in our design it was possible to include enough fan-in and fan-out memory requirements per core as established by those models and therefore no additional "routing core" would be allocated.
Hence, with the proposed architecture the scaling is only linear with the CNN size and so the advantage for larger networks, as required by increasingly complex tasks, is obvious since no additional "routing" cores are needed.

\section{Discussion}
\label{sec:discussion}

One of the most appealing features of neuromorphic processors is their ability to implement massively parallel computing architectures with memory and computation co-localized within their computational nodes. In addition to enabling ultra low-power data-driven processing, this feature allows for the construction of scalable systems for implementing very large scale neural networks, without running into the von Neumann bottleneck problem~\cite{Backus78,Indiveri_Liu15}, and in other data communication bottleneck problems typical of centralized systems. 
However, distributed memory/computing systems come at the price of non-trivial routing overhead. Neuromorphic solutions that have optimized this overhead and, with it, the communication bandwidth, have sacrificed flexibility and programmability~\cite{Benjamin_etal14}. On the other hand, solutions that have maximized network configurability, have sacrificed silicon real-estate for on-chip memory~\cite{Merolla_etal14a}, or resorted to using external memory banks~\cite{Furber_etal14} (and therefore eliminated the advantage of having memory and computation co-localized). \revised{rev1-5}{In this work we presented a scalability study for neuromorphic systems that allows}{} network configuration programmability and optimization of area and power consumption at the same time. This is especially important \revised{cnn}{for the configuration of different classes of sensory-processing neural networks that are not restricted to classical \acp{CNN} and \acp{DNN} architectures, and that can be used to process real-time streaming data \emph{on the fly}, without having to resort to saving input data in external memory structures (be it \ac{DRAM} or other forms of solid-state storage.}{in \acp{CNN} and \acp{DNN} because of the dependence of their performance on their actual structure, layer by layer.}

While there have been ad-hoc solutions proposed in the literature for utilizing off-chip memory more efficiently in neuromorphic systems~\cite{Davies_etal12}, there is no consensus around a systematic approach that would explicitly allow to trade-off flexibility and memory to meet the requirements of specific applications. The approach we proposed allows to choose an optimal trade-off point between network configuration flexibility and routing memory demands.
The trade-offs and design points we chose for the prototype device built to validate our approach were determined by the study of cortical networks in mammalian brains~\cite{Binzegger_etal04,Douglas_Martin07,Markov_Kennedy13}. In particular, our solution profits from the use of strategies observed in anatomically realistic topologies found in biology, namely to express dense clusters of connectivity that are interconnected by few long range connections. These choices resulted in a novel architecture that incorporates distributed and heterogeneous memory structures, with a flexible routing scheme that can minimize the amount of memory required for a given network topology, and that supports the most common feed-forward and recurrent neural networks. The use of on-chip heterogeneous memory structures makes this architecture ideal for exploiting the emerging memory structures based on nano-scale resistive memories~\cite{Rajendran_etal13,Akinaga_Shima10,Indiveri_etal16}.

\section{Conclusions}
\label{sec:conclusions}

In this work we presented a  two-stage routing scheme which minimizes the memory requirements needed to implement scalable and reconfigurable spiking neural networks with bounded connectivity. We used this scheme to  solve analytically the trade-off between ``point-to-point'' connectivity, which increases memory budget in favor of connection specificity, and ``broadcasting'', which reduces memory budget by distributing the same signal across populations of neurons.
We presented \ac{QDI} circuits and building blocks for implementing this routing scheme in asynchronous \ac{VLSI} technology, and presented a prototype neuromorphic processor that  integrates such building blocks  together with mixed signal analog/digital neuron and synapse circuits.
We showed that even with the conservative 0.18\,um \ac{VLSI} technology used, the power consumption of the prototype neuromorphic processor chip fabricated is comparable to state-of-art digital solutions. To demonstrate the features of such architecture and validate the circuits presented, we tiled multiple neuromorphic processors together and configured the setup to implement a three layer \ac{CNN}. We applied the network to an event-based data-set and showed how the system can produce reliable accurate results, using extremely low power and latency figures. As scaling studies have demonstrated that such architecture can outperform the current state-of-the-art systems when implemented using a 28\,nm \ac{FDSOI} process~\cite{Qiao_Indiveri16}, we are confident that such approach can lead to the design of a new generation of neuromorphic processors for solving a wide range of practical applications that require real-time processing of event-based sensory signals, using ultra low-power, low latency, and compact systems.


\section*{Acknowledgments}
We are grateful to Rajit Manohar, at Yale University, for providing us the tutorials and tips on the design of asynchronous circuits, as well as CAD tools for the synthesis and verification of the asynchronous circuits. We thank our colleagues at the Institute of Neuroinformatics in Zurich for stimulating discussions. This work was supported by the EU ERC Grant ``neuroP'' (257219).

\bibliographystyle{IEEEtran}

\section*{Appendix}
\label{sec:apx}
\subsection{Routing memory minimization constraints}
\label{sec:rout-memory-minim}
Here we provide an intuitive explanation of why the two-stage routing scheme presented in Section~\ref{sec:optimized} minimizes memory usage.
First, consider a simple case in which two subgroups of $K$ neurons project only within their groups.
In this case the two groups correspond in fact to two separate networks and there is no need to store more than $K$ words (which we shall call tags herein) to identify the neurons and thus to implement the connectivity for this network.
Obviously having a larger network with more groups that share the same property of local connectivity doesn't affect $K$, hence $K$ is constant with the number of neurons in the network.
It is clear that adding one connection from one group to an other may potentially cause an address collision, i.e., two neurons with the same tag project onto the same population, and therefore more than $K$ tags are required in this case in order to implement different connectivity patterns for those neurons.
If few connections exist between different clusters, i.e., in the case of sparse long-range connectivity, the total number of tags still scales slowly with the number of neurons because the number of address collisions is low, furthermore it can be kept under control by tag re-assignment.
Although the \revised{rev1-6}{particular}{particolar} number of $K$ tags for a network depends on the actual specific connectivity, its average value at least for a given class of networks scales nicely with network size for the above reasons.
An analogous situation appears if the neurons in one group never project back to the same group, as for example in a feed-forward network.
These are deliberately oversimplified cases but in many scenarios neural networks do show clustering and grouping properties.
It is clear that in such situations a $K$-tag based scheme for implementing the connectivity can be harnessed to gain in memory efficiency over a traditional scheme that requires to store $N$ different identities and their connectivity.
Next,we show a more systematic derivation of the theoretical constraints for optimizing digital memory in neuromorphic hardware implementations of clustered neural networks.

The optimal design point must satisfy the following requirements:
\begin{enumerate}
\item [1)] $F\geq M^*$
\item [2)] $C\geq M^*$
\end{enumerate}

If the first requirement is not met and $M^*>F$:
\begin{eqnarray*}\label{eq:first-constraint}
\sqrt{\frac{F}{\alpha}  {\frac{\log_2(\alpha N)}{\log_2(\alpha C)}}} & > & F\\
 \overset{\alpha=1}{\Rightarrow}  N^{1/F}>C 
\end{eqnarray*}

Therefore, $M^*$ is a valid design point if the cluster size meet this condition: $C \ge N^{1/F}$. This is a very safe
constraint: for example even when the total neuron count is in the $10^{10}$ range, a fan-out as small as $10$ would require a cluster
size of $C\ge 10$ to be able to have an optimal choice of $M^{*}$. Since typical fan-out values are actually in the
$10^3$--$10^4$ range, this requirement imposes very few constraints on the cluster size. The total number of neurons $N$ would have to be
larger than $10^{10^3}$ before the right hand side of the constraint would be $10$ or larger.\\
The second requirement indicates that the cluster size must be greater than a minimum size ($C\geq M^*$) to support the anticipated fan-out.
\begin{eqnarray*}\label{eq:second-constraint}
C & \geq & \sqrt{\frac{F}{\alpha} \frac{\log_2(\alpha N)}{\log_2(\alpha C)}}%
\end{eqnarray*}
which leads to:
\begin{eqnarray*}
\qquad \sqrt{F\log_2 \alpha N} & \leq & \sqrt{\alpha} C\sqrt{\log_2 \alpha C}\\
\qquad 
C\sqrt{\log_2(C)} & \geq & \sqrt{F\log_2(N)} \qquad\hbox{for $\alpha=1$}\\
\end{eqnarray*}
This constraint is much more restrictive than the first one. For
example, if we take typical values of $F=5000$, and  $N=10^{10}$,
then clusters need to be  $C \ge 152$.

Conversely, if we pick a cluster size $C=256$ with $\alpha=1$ (i.e., with 256 tags), then the optimal value of $M$ would be $M^*=144$.
In this case, the network requires a first-level fan-out of 35, followed by a second cluster-level fan-out of 144 for a \revised{rev1-7}{}{total maximum} fan-out of $5040$ an the storage per neuron would be 424.26$\sqrt{\log_2 N}$ bits.

\subsection{Communicating Hardware Process (CHP)}
\label{sec:comm-hardw-proc}

Here is a list of the most common \ac{CHP} commands that cover the design of all the blocks presented in this paper:
\begin{itemize}
\item Send: X$!$e means send the value of e over channel X.
\item Receive: Y $?$v means receive a value over channel Y and
store it in variable v.
\item Probe: The boolean expression X is true \revised{rev1-8}{if}{iff} a communication
over channel X can complete without suspending.
\item Sequential Composition: S; T
\item Parallel Composition: S ' T or S, T
\item Assignment: $a := b$. This statement means “assign the value
of b to a.” We also write a$\uparrow$ for $a := true$, and a$\downarrow$ for
$a := false$.
\item Selection: $[$G1 $\rightarrow$ S1 [] ... [] Gn $\rightarrow$ Sn], where Gi’s
are boolean expressions (guards) and Si’s are program parts.
The execution of this command corresponds to waiting until
one of the guards is true, and then executing one of the
statements with a true guard. The notation $[G]$ is shorthand
for [G $\rightarrow$ skip], and denotes waiting for the predicate
G to become true. If the guards are not mutually exclusive,
we use the vertical bar “|” instead of “$[]$.”
\item Repetition: *[G1 $\rightarrow$ S1 [] ... [] Gn $\rightarrow$ Sn]. The execution
of this command corresponds to choosing one of the true
guards and executing the corresponding statement, repeating
this until all guards evaluate to false. The notation *[S] is
short-hand for *[true $\rightarrow$ S].

\end{itemize}

\subsection{CHP, HSE, and PR examples or asynchronous circuits used in the routing scheme}
\label{sec:chp-hse-pr}

\begin{lstlisting}[language=chp, caption=merge, escapeinside={(*}{*)}, mathescape=true, breaklines=true, basicstyle=\footnotesize]
chp{
     *[[$ \overline{in1} $ -> in1?s; out!s
        | $ \overline{in2} $ -> in2?s; out!s
      ]]
     }

hse{
     *[[ in1arb_out -> x+; 
      out.b[i].d[j] := in1.b[i].d[j]; 
      in1.a+; en1-;
       ([out.a]; x-; out.b[i].d[j]-; 
        [~out.a]), (~in1arb_out]; 
         in1.a-); en1+
        []
         in2arb_out -> x+; 
         out.b[i].d[j] := in2.b[i].d[j];
          in2.a+; en2-;
        ([out.a]; x-; out.b[i].d[j]-;
         [~out.a]), (~in2arb_out];
          in2.a-); en2+
        ]]

        ||

         *[in1.v & en1 -> in1arb_in+;
          [~in1.v & ~en1]; in1arb-]

             ||

             *[in2.v & en2 -> in2arb_in+;
              [~in2.v & ~en2]; in2arb-]
              
              ||
           
              *[[in1arb_in -> in1arb_out+; 
                [~in1arb_in]; in1arb_out
                - |in2arb_in -> in2arb_out+;
                [~in2arb_in]; in2arb_out-
               ]]
      }
   
      \end{lstlisting}

\begin{lstlisting}[language=chp, caption=controlled-split, escapeinside={(*}{*)}, mathescape=true, basicstyle=\footnotesize]
chp{
        *[[v(in )]; [ctrl0 -> out0==
                    [] ctrl1 -> out11==];
                    [n(in ) ^ n(ctrl)]; 
                    out0=!, out1=!]
    }
\end{lstlisting}
      
\begin{lstlisting}[language=chp, caption=buffer, escapeinside={(*}{*)}, mathescape=true, basicstyle=\footnotesize]
chp{
        *[ IN?x ; OUT!x ] 
    }
\end{lstlisting}

\subsection{Asynchronous Content-Addressable Memory (CAM) timing assumptions}
\label{sec:timing-assumptions}
\begin{figure}
\centering
\includegraphics[width=0.4\textwidth]{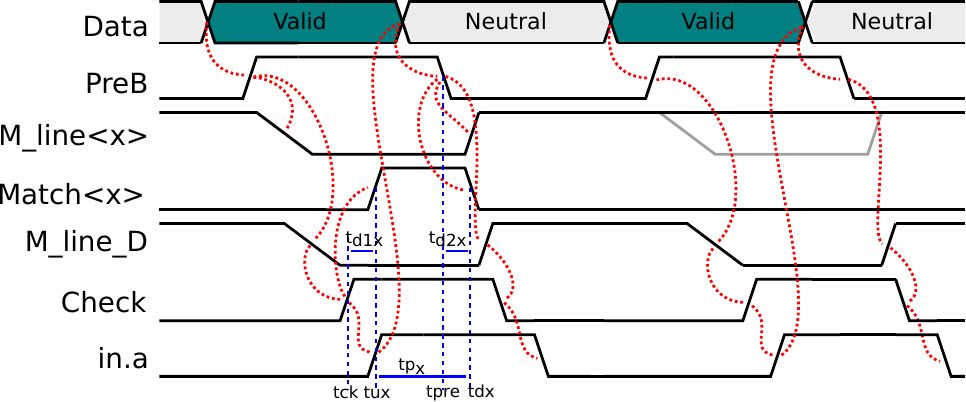}
\caption{CAM's 4-phase hand-shaking protocol.}
\label{fig:cam_handshaking}
\end{figure}

In order to guarantee the correct handshaking communication between the local router (R1) and the CAM array, it is necessary to make appropriate timing assumptions. 
Here we describe the assumptions that were made in the design of the asynchronous \ac{CAM} array. 
Initially, without the presence of input data, the output of Event Buffer ($EB$) is in neutral state and all the search lines are low: $SL\left\langle9:0\right\rangle=0$ and $SLB\left\langle9:0\right\rangle=0$.
Therefore the Validity Check (VC) block, placed at the top-right-hand side of the array in the layout (see also Fig.~\ref{fig:cam}, sets $PreB$ to 0 and consequently, all the MLs are pre-charged to high. Upon receiving the events, the buffers start broadcasting and driving the data lines in entire array. The $VC$, the last element in the array to receive the valid data, eventually asserts $PreB$ to high.
This signal is in turn used to enable the comparison between the input (presented on data lines) and the CAM words content by sending the enable signal across the whole array from top-left to bottom-right.
The duplicate CAM cell CAM\_D, placed at the opposite corner to the $VC$ block (at the bottom-right-hand side of the CAM array), is assumed to be the last one to get the enable signal $PreB$.
This duplicate cell is designed to produce a $miss$ signal with the worst case (that is when only one bit is a miss in its 10-bit CAM word) for any data presented on search lines. 
Once its ML M\_line\_D is discharged, which guarantees that the comparison between input data lines and all CAMs words have been finished. At this point $Check$ signal is being driven to high and is sent across the array. This signal then lets the CAM words with active MLs produce a $match (hit)$ signal.  The Check signal eventually reaches the $EB$ block as the acknowledgment from the array.
The $EB$ circuits then de-asserts data and sets data lines to their neutral state, which lowers the $PreB$. Again, the $PreB$ signal is sent from top-left to bottom-right to reset all the $Match$ lines to low as well as to reset all MLs. As soon as the ML of the dummy CAM\_D, word is lowered, the $Check$ signal is de-asserted and the handshaking is completed at this point.
Although the local communication is not optimized for speed, it is sufficient to cover all the neurons in one core.
The way we implemented the timing assumption guarantees small mismatch of pulse width of $Match$ signals generated by different CAM words in the array.
The pulse width is a critical parameter for the analog neural computations. It is ensures that the mismatch between width of $Match$ pulses generated in different synapses across the core is as small as possible.
Assuming $PreB$ and $Check$ signals drive similar size load capacitances, The same buffer is used for sending these signals from top left to bottom right-hand side of the array. For a particular CAM word, CAMx, $t_{d1x}$, the delay between transmitting the ``Check'' signal and Match<x>, and $t_{d2x}$, the delay between $PreB$ and Match<x> are about the same(see Fig.~\ref{fig:cam_handshaking}).
The pulse width of a particular CAMx is $t_{dx}t_{ux} =(t_{\mathrm{pre}} +t_{d2x})-(t_{ck} +t_{d1x})$.
And for $t_{d1x}$ and $t_{d2x}$, the pulse width can be approximated to $t_{\mathrm{pre}}-t_{ck}$ which is \revised{rev1-10}{not related to the physical position of the CAM.}{no more related.}

In designing the asynchronous \ac{CAM} array we made worst-case scenario timing assumptions to ensure the correct operation of the circuits.
Lets consider the case in which the dual-rail data from the output of Event Buffer (EB) of Fig.~\ref{fig:cam} is neutral with all search lines: SL<9:0>=0 and SLB<9:0>=0.
The Validity Check (VC) block at the top-right-hand side of the array in Fig.~\ref{fig:cam} checks the data state and assigns PreB to 0 for the neutral state.
All MLs are consequently pre-charged to high.

Once EB gets events from the previous stage, i.e., from the core router R1, it pushes the new data to dual-rail data lines and broadcasts it to the whole array through broadcasting buffers.
After the searching lines have successfully set-up the new data, VC is assumed to be the last one in this array to get the valid data.
This is in turn used to enable the \ac{CAM}s comparisons by transmitting the enable signal to the whole core from top-left to bottom-right.
The duplicate \ac{CAM} cell \ac{CAM}\_D, placed at the bottom-right-hand side of the \ac{CAM} array, is assumed to be the last one to get the enable signal PreB for comparing.
This duplicate cell is designed to always get a miss with the worst case (only one bit is a miss in its 10-bit \ac{CAM} word for discharging the ML) for any data presented on search lines.
Once its ML M\_line\_D is discharged, which guarantees that all \ac{CAM}s in this array have finished comparing, an inverted signal Check is asserted to high and sent to left-top of the array.
Then it is buffered to the whole array from left-top to right-bottom.
All MLs in this array will apply an AND operation with Check, PreB and ML, and \ac{CAM} words with match state will get a high Match.
The Check signal will finally arrive at EB as the acknowledge signal of its valid output data.
The EB will then de-assert the data
and set the data lines to neutral again, which will cause the PreB from VC to be set to 0.
Again, the high to low transition of PreB is transmitted from top-left to bottom-right of the layout, to reset all high Match to low as well as to reset all MLs.
As soon as this happens the ML of the dummy cell \ac{CAM}\_D (which has been positioned physically as the last in the array), is reset again and the signal Check is set to low, to finish the handshaking.
Assuming that PreB and Check have similar load capacitances in the whole array and that the same buffer is used for transmitting from top left to bottom right-hand side of the array, then for a particular \ac{CAM} word \ac{CAM}x in a match state, the delay $t_{d1x}$ between transmits of Check and Match<x> and the delay $t_{d2x}$ between PreB and Match<x> will be similar.



Another critical circuit is the one responsible for extending input pulses. Since that pulse width is also one critical parameter of analog computing, the pulse width circuit has to minimize the effct of mismatch, such that pulses generated in different synapses across the core should be as similar as possible.
Figure~\ref{fig:cam_handshaking} shows the sequence of operations from the events received at the input of each to the signals routed to the synapses in the core. 
At the very beginning, dual-rail data from output of Event Buffer (EB) is "neutral" with all search lines, $SL\left\langle9:0\right\rangle=0$ and $SLB\left\langle9:0\right\rangle=0$.
The Validity Check (VC) block at the right-top of the array will check the data state and assign $PreB$ to "0" for the "neutral" state.
All MLs are consequently pre-charged to "high".
Once EB gets events from the previous stage, i.e.\ the local router R1, it will push the new data to dual-rail data lines and broadcast it to the whole array through broadcasting buffers.
After searching lines have successfully set-up new data, VC is assumed to be the last one in this array to get the valid data, check dual-rail data presented on search lines and assert $PreB$ to "1" for the "valid" state.
This is in turn used to enable the \ac{CAM}s comparisons by transmitting the ``enable'' signal the whole core from left-top to right-bottom.
The duplicate \ac{CAM} cell $CAM\_D$ placed at the right-bottom of the \ac{CAM} array is assumed to be the last one to get the enable signal $PreB$ for comparing. 
This duplicate cell is designed to always get "miss" with the worse case (only one bit is miss in its 10-bit \ac{CAM} word for discharging the ML) for any data presenting on search lines.
Once its ML $M\_line\_D$ is discharged which can guarantee that all \ac{CAM}s in this array have finished comparing, a inverted signal $Check$ is asserted to "high" and sent to left-top of the array and then buffered to the whole array from left-top to right-bottom.
All MLs in this array will do \revised{rev1-9}{$AND$}{$and$} operation with $Check$, $PreB$ and $ML$, \ac{CAM} words with "match" state will get "high" $Match$.
The $Check$ signal will finally arrive at EB as the acknowledge signal of its valid output data.
The EB will then dessert data and set data lines to "neutral" again, which will cause the $PreB$ from VC to be "0".
Again, the "jumping from high to low" of $PreB$ is still transmitted from left-top to right-bottom to reset all "high" $Match$ to "low" as well as reset all MLs.
As soon as the ML of duplicate cell \ac{CAM}\_D which is assumed to be the last one in the array has been reset again, $Check$ will go low and set in.a to low to finish the handshaking.

The local communication is not optimized for speed, however it minimizes the mismatch of the pulse width generated by different \ac{CAM} words with {\em match}\/ states distributed in the whole array. Due to the small number of neurons in each core, the speed of operations is already sufficient with this scheme. 
Assuming $PreB$ and $Check$ have similar load capacitances in the whole array and that the same buffer is used for transmitting from top-left to right-bottom of the array, then for a particular \ac{CAM} word $CAM_{x}$ in a {\em match}\/ state, the delay $t_{d1x}$ between transmits of $Check$ and $Match<x>$ and the delay $t_{d2x}$ between $PreB$ and $Match<x>$ will be similar (see Fig~\ref{fig:cam_handshaking}).
The pulse weight of a particular $CAM_{x}$ is $t_{dx} - t_{ux} = (t_{pre} + t_{d2x}) - (t_{ck} + t_{d1x})$.
For similar $t_{d1x}$ and $t_{d2x}$, the pulse weight can be approximated to $t_{pre} - t_{ck}$ which is no more related to the physical position of the \ac{CAM}.

\end{document}